%% file: HyperGraph-arXiv-main.tex
\documentclass[10pt,journal,compsoc]{IEEEtran}

\ifCLASSOPTIONcompsoc
  \usepackage[nocompress]{cite}
\else
  \usepackage{cite}
\fi

\ifCLASSINFOpdf
  \usepackage[pdftex]{graphicx}
  \graphicspath{{./figs/}}
  \DeclareGraphicsExtensions{.pdf,.jpeg,.png}
\else
   \usepackage[dvips]{graphicx}
   \graphicspath{{./figs/}}
   \DeclareGraphicsExtensions{.eps}
\fi

\usepackage{amsmath}
\usepackage{algorithmic}
\usepackage{array}
\usepackage{url}

\usepackage{microtype}                
\PassOptionsToPackage{warn}{textcomp}  
\usepackage{textcomp}                  
\usepackage{mathptmx}                  
         
\usepackage{cite}                      
\usepackage{tabu}                     
\usepackage{booktabs}

\usepackage{amssymb}
\usepackage{hyphenat}
\usepackage{mathrsfs}
\usepackage{graphicx}
\usepackage{subcaption}
\usepackage{microtype}                 
\usepackage{textcomp}                 
\usepackage{mathptmx}                   
\usepackage{cite}                      
\usepackage{tabu}                      
\usepackage{booktabs}                 
\usepackage{amsthm} 
\usepackage{amsfonts}
\usepackage{bm}

\usepackage{xcolor}
\usepackage{hyperref}

\newcommand {\mm}[1] {\ifmmode{#1}\else{\mbox{\(#1\)}}\fi}

\newcommand{\Rspace}        {\mm{\mathbb{R}}}

\newcommand{\Bcal}        {\mm{\mathcal B}}

\theoremstyle{definition}

\newtheorem{definition}{Definition}[section]

\newcommand{\para}[1]        {\vspace{2mm}\noindent{\textbf{#1}}}

\newcommand{\etal}{\textit{et al.}}
\newcommand{\wrt}{\textit{w.r.t.}}

\newcommand{\lesmis}{Les Mis\'{e}rables}

\usepackage{mathptmx} 
         
\DeclareMathAlphabet{\mathcal}{OMS}{cmsy}{m}{n}
\usepackage{amsfonts}
\usepackage{amssymb}
\usepackage{amsmath}
\usepackage{lipsum,multicol}
\usepackage{tabularx}
\usepackage{multirow}
\usepackage{gensymb}

\begin{document}

\title{Topological Simplifications of Hypergraphs}

\author{Youjia Zhou, Archit Rathore, Emilie Purvine, Bei Wang
\IEEEcompsocitemizethanks{\IEEEcompsocthanksitem Youjia Zhou, Archit Rathore, Bei Wang are with Scientific Computing \& Imaging (SCI) Institute, University of Utah, Salt Lake City, UT, 84112.
E-mails: \{zhou325, archit, beiwang\}@sci.utah.edu.
\IEEEcompsocthanksitem Emilie Purvine is with Pacific Northwest National Laboratory, Seattle, WA, 98109. 
E-mail: Emilie.Purvine@pnnl.gov.}}

\markboth{~}%
{Zhou \MakeLowercase{\textit{et al.}}: Hypergraph Visualization via a Metric Space Viewpoint}

\IEEEtitleabstractindextext{%
\begin{abstract}
\input{sec-abstract.tex}

\end{abstract}

}

\maketitle

\IEEEdisplaynontitleabstractindextext
\IEEEpeerreviewmaketitle

\input{sec-introduction.tex}

\input{sec-related-work.tex}

\input{sec-background.tex}

\input{sec-methods-simplify.tex}

\input{sec-vis.tex}

\input{sec-results.tex}
\input{sec-evaluation.tex}

\input{sec-discussion.tex}

\section*{Acknowledgments}
This project was partially supported by NSF IIS 1513616 and 
DOE DE-SC0021015. 

\input{HyperGraph-arXiv-main.bbl}

\input{sec-bio.tex}

\end{document}

%% file: sec-abstract.tex
We study hypergraph visualization via its topological simplification. 
We explore both vertex simplification and hyperedge simplification of hypergraphs using tools from topological data analysis. 
In particular, we transform a hypergraph to its graph representations known as the line graph and clique expansion. 
A topological simplification of such a graph representation induces a simplification of the hypergraph. 
In simplifying a hypergraph, we allow vertices to be combined if they belong to almost the same set of hyperedges, and hyperedges to be merged if they share almost the same set of vertices. 
Our proposed approaches are general, mathematically justifiable, and they put vertex simplification and hyperedge simplification in a unifying framework.

%% file: sec-introduction.tex
\section{Introduction}
\label{sec:introduction}

\IEEEPARstart{D}{ata} that capture multiway relationships within a group of entities are ubiquitous in science and engineering. In social networks,  apart from pairwise ``likes" and friendships, people form multi-way  groups or clubs based on common interests. 
In computer networking, multiple IP addresses that resolve to the same domain name (e.g., \texttt{www.google.com}) form a group relationship~\cite{JoslynAksoyArendt2020}. 
In biological applications, collections of proteins comprise pathways that lead to a product or a change in a cell, and groups of genes make up ontology terms and contribute toward a shared molecular function, cellular component, or biological process~\cite{AshburnerBallBlake2000,GeneOntologyConsortium2019}. 

\begin{figure}[!ht]
\centering
\includegraphics[width=0.7\columnwidth]{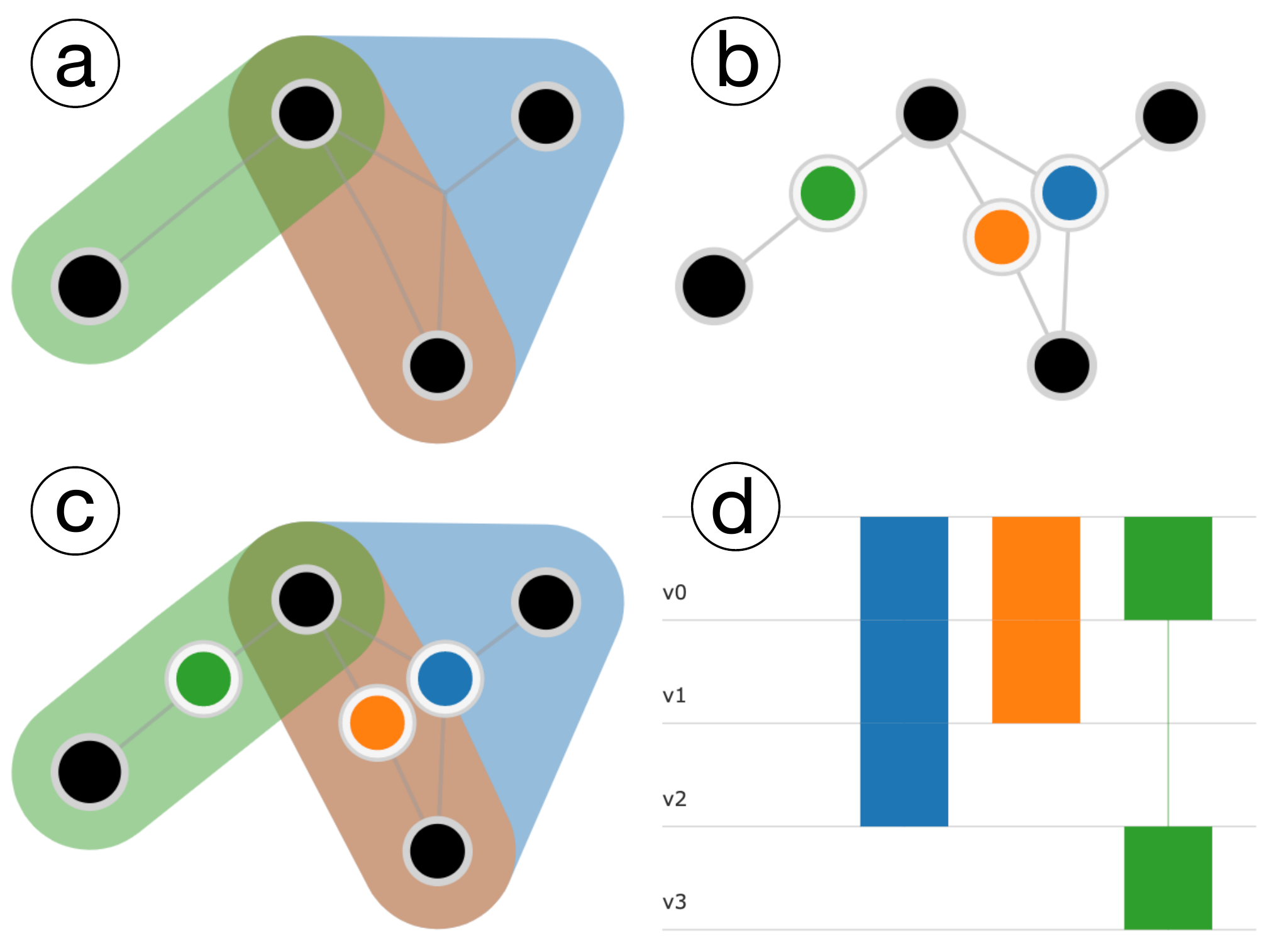}
\caption{Various visual encodings of a hypergraph. (a) Venn diagram based  visualization: black nodes are vertices, colored convex hulls are hyperedges. (b) Bipartite graph based visualization with force-directed layout: each colored node represents a hyperedge, which connects with all its vertices in black. (c) Hybrid visualization by superimposing Venn diagram with bipartite graph based visual encodings. (d) Rainbow box based visualization: each row is a vertex, each column is a hyperedge.}
\label{fig:visual-encoding}
\end{figure}

In all of these cases, exploration of the data can help discover interesting patterns, subsets, and entities.
Graphs (or networks) are a central way to model data that come in the form of pairwise (or binary) relationships. 
However, graphs cannot natively represent multiway relationships without moving to bipartite structures or employing reification strategies. 
Instead, hypergraphs provide a way to capture these multiway interactions. 
A \emph{hypergraph}, $H = (V,E)$, consists of a set of vertices, $V = \{v_1,\cdots ,v_n\}$, together with a collection of \emph{hyperedges}, $E = \{e_1,\cdots ,e_m\}$, each of which is a subset of vertices $e_i \subseteq V$. 
For instance, \autoref{fig:visual-encoding}a gives a hypergraph with 4 vertices and 3 hyperedges.

Visualization can be a useful tool to explore data modeled as hypergraphs. 
There are various visual encodings for a hypergraph. 
A Venn diagram based visualization (\autoref{fig:visual-encoding}a) places vertices on the plane and represents hyperedges as convex hulls of the vertices.  
A bipartite graph based visualization (\autoref{fig:visual-encoding}b) considers the set of hyperedges and the set of vertices as the two disjoint sets of a bipartite graph.  
A hybrid visualization combines both Venn diagram with bipartite graph based visualizations (\autoref{fig:visual-encoding}c).  
And a rainbow box based visualization treats hyperedges (columns) as boxes (\autoref{fig:visual-encoding}d).   

\begin{figure*}
    \centering
    \includegraphics[width=0.98\textwidth]{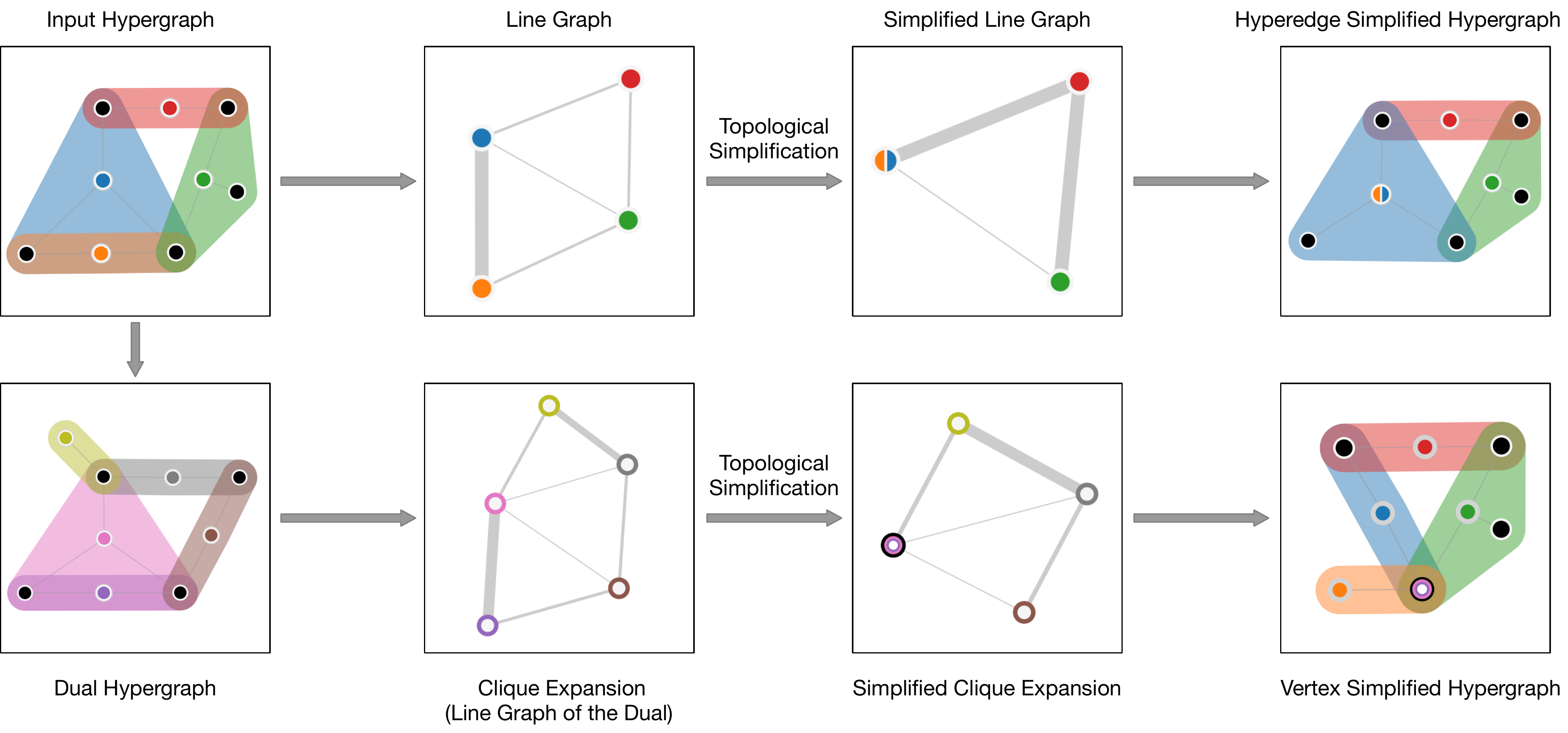}
    \caption{An overview of topological simplification of hypergraphs. Top: hyperedge simplification. Bottom: vertex simplification.}
    \label{fig:overview}
\end{figure*}

\begin{figure}[h]
    \centering
    \includegraphics[width=0.99\columnwidth]{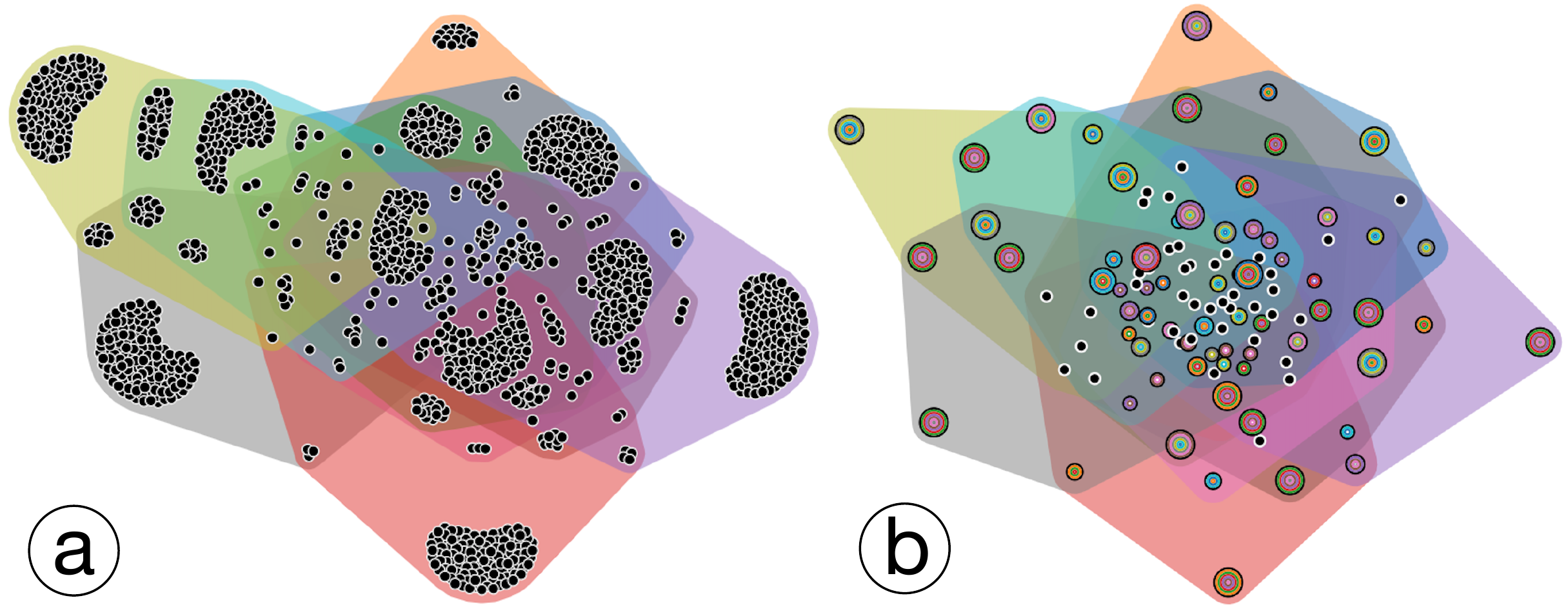}
    \caption{(a) A biological pathway hypergraph with $|V|=1,316$ and $|E|=10$.   
Black nodes are vertices, convex hulls are hyperedges. 
(b) The simplified hypergraph after vertex collapse.}
    \label{fig:s25-e10}
\end{figure}

Visualizing large graphs remains challenging as naive visualization often produces ``hairballs" of little information content due to visual clutter. 
Such a problem is further compounded when dealing with hypergraphs, even ones with a moderate number of vertices and a small number of hyperedges. 
As the number of vertices grows and hyperedges become more interconnected, it becomes increasingly difficult to turn a naive visualization into insights. 
\autoref{fig:s25-e10}a illustrates a naive hypergraph visualization where vertices represent genes and hyperedges consist of pathways from the Hallmarks collection within the Molecular Signatures Database (MSigDB)~\cite{LiberzonBirgerThorvaldsdottir2015,SubramanianTamayoMootha2005}. 
While vertices on the periphery are shown to belong to a single hyperedge, edge memberships of those vertices closer to the center are more difficult to interpret.

\begin{figure}[!ht]
    \centering
    \includegraphics[width=0.76\columnwidth]{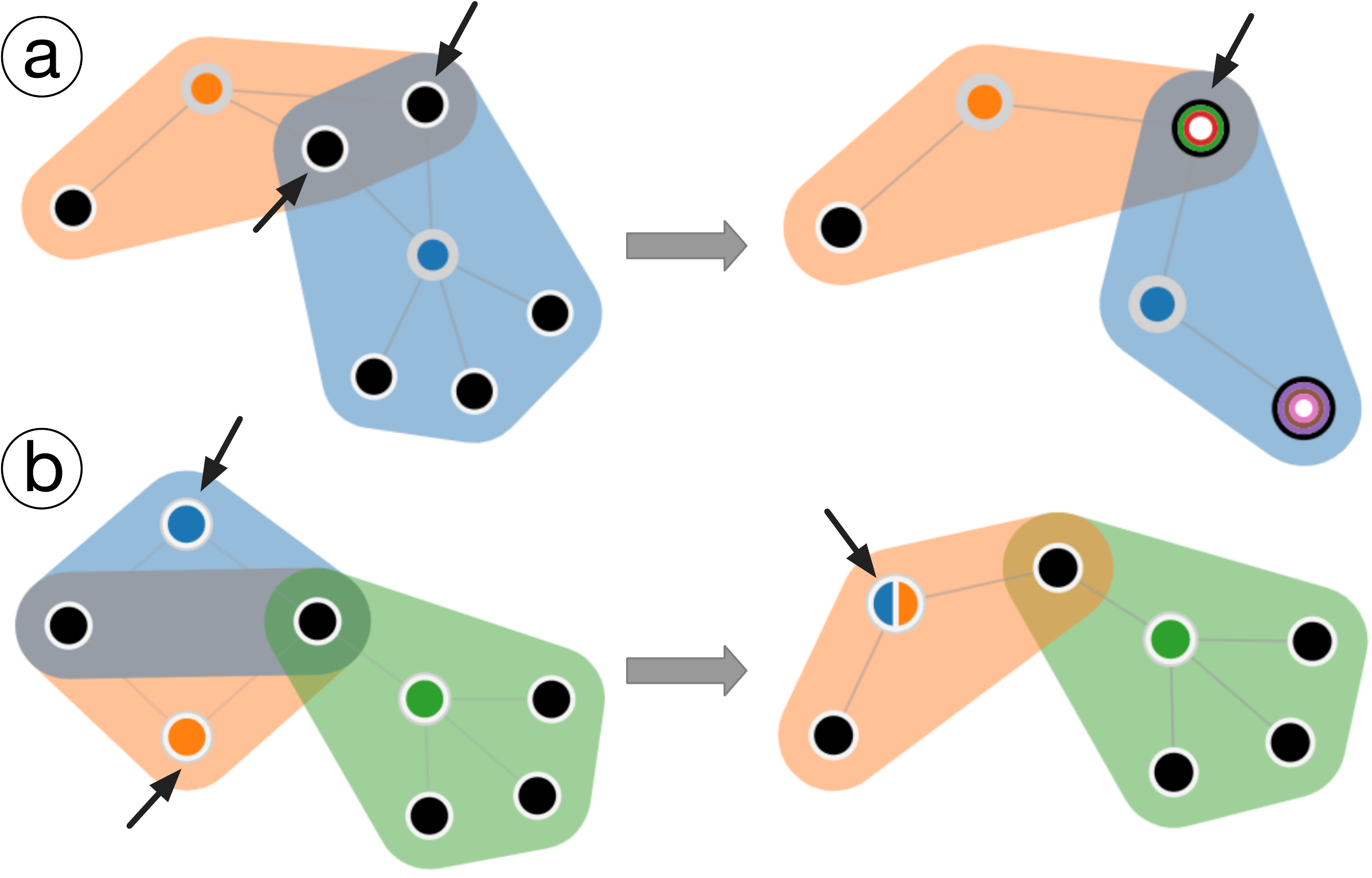}
    \vspace{-2mm}
    \caption{A vertex collapse (a) and a hyperedge collapse (b).}
    \label{fig:collapse}
\end{figure}

To reduce visual clutter, we might want to apply \emph{vertex collapse} and \emph{hyperedge collapse} to reduce the size of the hypergraph.
As illustrated in~\autoref{fig:collapse}, vertex collapse combines vertices that belong to exactly the same set of hyperedges into a single ``super-vertex" (visualized by  concentric ring glyph), while hyperedge collapse merges hyperedges that share exactly the same set of vertices into a ``super-edge'' (visualized by a pie-chart glyph).  
See~\autoref{fig:s25-e10}b for the simplified biological pathway hypergraph after vertex collapse. 

In this paper, we relax the notions of vertex collapse and hyperedge collapse by allowing vertices to be combined if they belong to \emph{almost the same} set of hyperedges, and hyperedges to be merged if they share \emph{almost the same} set of vertices. 
The former is referred to as the \emph{vertex simplification} (or approximate vertex collapse), and the latter is referred to as the \emph{hyperedge simplification} (or approximate hyperedge collapse).
Using tools from topological data analysis, in particular, barcodes that capture the topology of hypergraphs, we perform topological simplification of hypergraphs. 
Our approach is \emph{general} as it generalizes vertex and hyperedge collapses to their approximate versions. 
As we simplify a hypergraph in a way that removes topological noise as determined by its barcode, our approach is also \emph{mathematically justified} by the stability of barcodes~\cite{Cohen-SteinerEdelsbrunnerHarer2007}.  

Our pipeline is illustrated in~\autoref{fig:overview}. To enable hyperedge simplification (\autoref{fig:overview} top), we first convert a hypergraph into a graph representation called the (weighted) line graph. 
A line graph captures the similarities among hyperedges. 
We then perform a topological simplification of the line graph. 
The simplified line graph induces a hyperedge simplification of the input hypergraph. 
On the other hand, to enable vertex simplification (\autoref{fig:overview} bottom), we first consider a (weighted) clique expansion of an input hypergraph. 
We then perform a topological simplification of the clique expansion. 
A clique expansion captures the similarities among vertices; it is known to be the line graph of the dual of a hypergraph. 
The simplified clique expansion in turn induces the vertex simplification of the input hypergraph. 
Using barcode-guided topological simplification, we formalize both vertex and hyperedge simplification in a \emph{unifying} way. 

We demonstrate the utility of our approach with real world examples. 
We provide an open source, interactive tool that includes modular and easily extendable implementation of our proposed algorithms.  
The tool allows users to explore the simplification framework by inputting their own datasets and applying vertex and hyperedge simplifications to gain insights from the data.
The tool is open source via GitHub:  
\texttt{https://github.com/tdavislab/Hypergraph-Vis}.

%% file: sec-related-work.tex
\section{Related Work}
\label{sec:related-work}

We focus on visualization techniques relevant to hypergraphs. 
For graph visualization, see surveys on graph visualization for information visualization~\cite{HermanMelanconMarshall2000}, graph representations for scientific visualization~\cite{WangTao2017}, visual analysis of large graphs~\cite{LandesbergerKuijperSchreck2011}, dynamic graphs~\cite{BeckBurchDiehl2014}, and graph drawing~\cite{EadesTamassia1994}. 

M\"{a}kinen~\cite{Makinen1990} introduced two widely used approaches for drawing hypergraphs. 
In an \emph{edge-based} approach, hyperedges are drawn as smooth curves connecting their vertices.   
In a \emph{subset-based} approach, they are drawn as closed curves enclosing their vertices. 
For the edge-based approach, by mapping a hypergraph to a graph, hypergraph visualization could be considered as an extension of graph visualization. 
Arafat \etal~\cite{ArafatBressan2017} proposed four different ways of encoding a hypergraph as a graph, via complete-, star-, cycle-, and wheel-associated-graphs. 
Paquette \etal~\cite{PaquetteTokuyasu2011} considered a hypergraph as a bipartite graph, where hyperedges and vertices form two disjoint and independent sets. 
For the subset-based approach, hypergraph visualization is closely related to set visualization (see~\cite{AlsallakhMicallefAigner2016} for a survey), which goes back to Euler diagrams~\cite{Euler1761} and its more restrictive form, the Venn diagrams. 
Kritz and Perlin~\cite{KritzPerlin1994} proposed the QUAD scheme that resembled a matrix encoding of set relations: each hyperedge is a column represented by a rectangle and each vertex is a point along a particular row. 
Simonetto \etal~\cite{SimonettoAuberArchambault2009, Simonetto2011} introduced an automatic generation of Euler-like diagrams (\emph{EulerView}) for any collection of sets and their intersections. 
Many recent works focused on representing sets in more efficient ways, including LineSets~\cite{AlperRicheRamos2011}, BubbleSets~\cite{CollinsPennCarpendale2009}, MapSets~\cite{EfratHuKobourov2015}, UpSet~\cite{LexGehlenborgStrobelt2014}, and LinearDiagrams~\cite{RodgersStapletonChapman2015} etc. 
The rainbow box based visualization implemented in our tool is inspired by the work of Lamy~\cite{Lamy2019}, which visualized undirected graphs and symmetric square matrices by transforming them into overlapping sets, and visualized them with rainbow boxes. 
The HyperNetX Python package~\cite{HyperNetX} includes hypergraph visualization using an Euler diagram approach. It also includes the ability to perform \emph{exact} edge and vertex collapses (as opposed to \emph{approximate} collapses which this current paper explores). 
Collapsed vertices and hyperedges are visualized in HyperNetX as larger ``supervertices'' and thicker ``superedges''.

In terms of layouts, Valdivia \etal~\cite{ValdiviaBuonoPlaisant2021} introduced Parallel Aggregated Ordered Hypergraph (PAOH) as a hybrid of edge-based and subset-based (matrix) approach, which represents vertices as parallel horizontal bars and hyperedges as vertical lines, using dots to depict the connections to one or more vertices. 
Kerren and Jusufi~\cite{KerrenJusufi2013} introduced a radial layout, where vertices are evenly distributed on a circle, and the hyperedges are represented as arcs that enclose the circle.
Hypergraphs can be represented geometrically~\cite{BrandesCornelsenPampel2012,BuchinKreveldMeijer2009}, starting with Zykov~\cite{Zykov1974}. 
Gropp~\cite{Gropp1995} positioned the vertices in the plane such that those that form hyperedges are collinear in the plane.
Evans \etal~\cite{EvansRzazewskiSaeedi2019} used polygons to represent hyperedges in 3D to gain additional flexibility.

Evaluating hypergraph visualizations can be considered from a quantitative and a qualitative perspective. 
Many evaluation criteria for graph visualization are applicable for hypergraphs (e.g.~\cite{MeidianaHongEades2019,NguyenEadesHong2012}), including aesthetic criteria such as \emph{readability}~\cite{MeidianaHongEades2019} and \emph{faithfulness}~\cite{NguyenEadesHong2012}. 
M\"{a}kinen~\cite{Makinen1990} gave a set of desirable aesthetic properties for subset-based hypergraph visualization.  
Arafat \etal~\cite{ArafatBressan2017} perfected these properties by introducing quantitative metrics such as \emph{concavity}, \emph{planarity}, and \emph{coverage} metrics. 
In our work, we evaluate the quality of hypergraph visualizations after simplification using four different aesthetic criteria. The first criterion evaluates the Venn diagram based visualization, which is to minimize the approximate number of contour intersections. The remaining three criteria evaluate the bipartite graph based visualization, which aim to minimize the number of edge crossings~\cite{Purchase2002}, to minimize the normalized edge length variation~\cite{KwonCrnovrsaninMa2017}, and to maximize the minimum angle between edges out from a node~\cite{Purchase2002}, respectively. 

For graph simplification, a number of works focused on algorithmic developments~\cite{purohit2014fast, ShinGhotingKim2019, beg2018scalable} and visualization~\cite{ShenMaEliassiRad2006, LeeJoKo2020, KoutraKangVreeken2014, ShahKoutraZou2015, DunneShneiderman2013}. 
In particular, Suh \etal~\cite{SuhHajijWang2019} proposed a topology-based graph simplification tool, which enables contraction of edges whose weight is below a user-specified threshold.
There are only a few research works on hypergraph simplification. 
Lemonnier \etal~\cite{LemonnierWeteringKissinger2020} proposed a hypergraph simplification approach using notions from quantum physics. 
To the best of our knowledge, we are the first to use topological profiles to guide the hypergraph simplification process for visual exploration.

Finally, many research efforts have focused on visualizing large graphs with advanced hardwares such as GPUs, e.g., Graphistry (\texttt{https://github.com/graphistry/}), see~\cite{LandesbergerKuijperSchreck2011} for surveys. Few works focus on large hypergraph visualization. 
We consider these scalable visualization approaches to be  tangential to hypergraph simplification. 

In this paper, we focus on increasing the readability while preserving as much as possible information faithfulness of hypergraph data via simplification. 
We apply vertex simplification and edge simplification to reduce the size of the hypergraph while preserving its information for insight discovery. 
It is important to emphasize that our simplification applies to \emph{any} hypergraph visualization technique. We primarily use subset based approaches for hypergraph visualization, including Venn diagram, bipartite graph, and rainbow box based approaches (\autoref{fig:visual-encoding}). 
While other visual encodings are possible for hypergraphs, we choose these approaches since the first two approaches capture spatial relations among the hyperedges, while the third approach highlights overlaps among the hyperedges.

%% file: sec-background.tex
\section{Technical Background}
\label{sec:background}

We review the definitions of hypergraphs and several graph representations highly relevant to hypergraphs, namely, dual hypergraphs, line graphs, and clique expansions.  
To explore graphs, one might employ network science measures such as centrality, clustering coefficient, and connected components to discover entities of interest.  
In this paper, we work with analogous \emph{hypernetwork science}~\cite{aksoy2020hypernetwork} concepts for hypergraphs, namely, $s$-walks and $s$-connected components. 

\subsection{Hypergraphs, Line Graphs, and Clique Expansions}
\label{subsec:hypergraphs} 
\begin{definition}
A \emph{hypergraph} $H = (V, E)$ consists of a set of $n$ \emph{vertices}, $V = \{v_1, \cdots, v_n\}$, and a family of $m$ \emph{hyperedges}, $E = \{e_1, \cdots, e_m\}$; where $e_i \subseteq V$ for $i \in [m]$. 
\end{definition} 

\begin{figure}[!ht]
  \centering
\includegraphics[width=0.7\columnwidth]{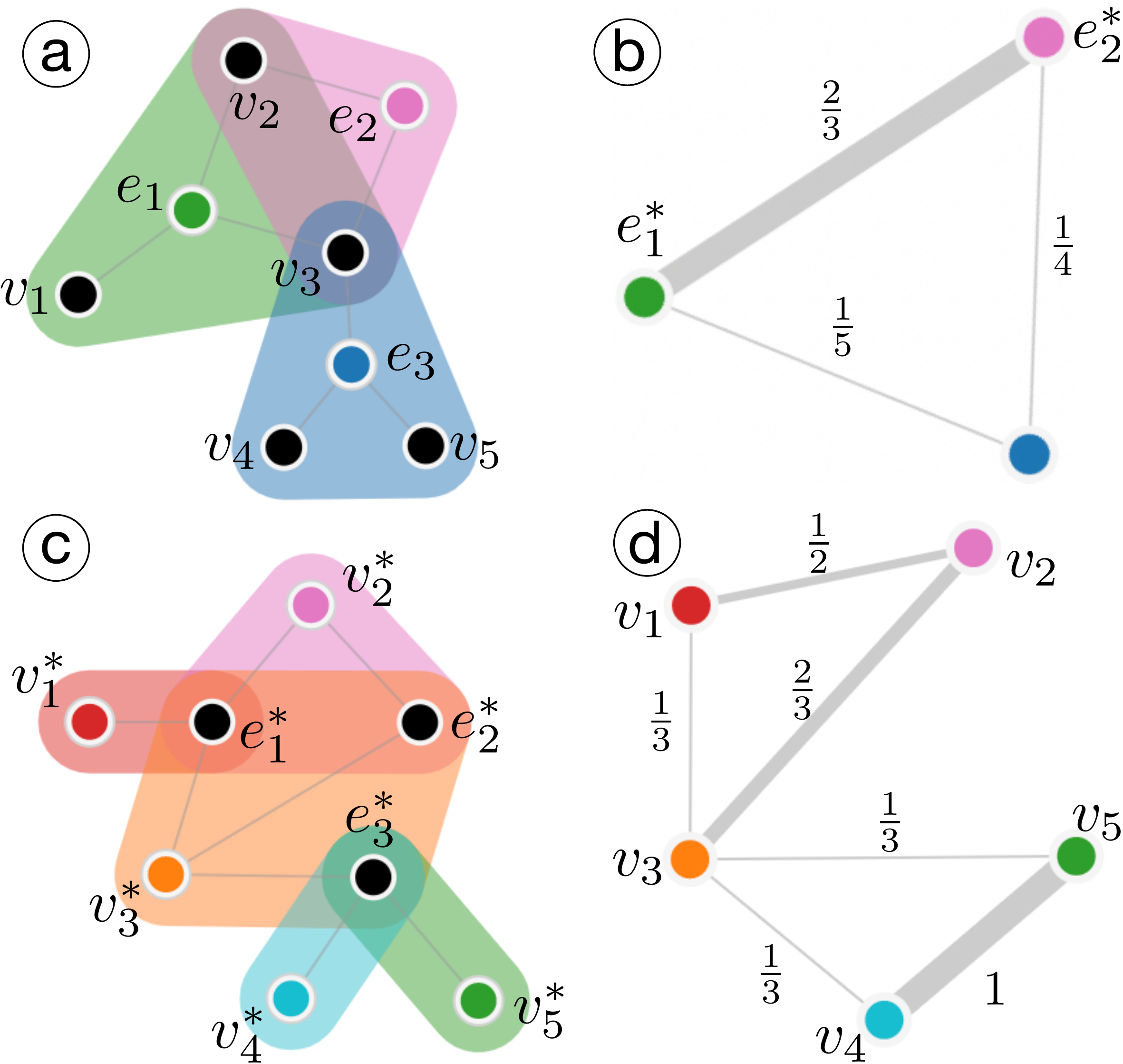} 
    \caption{(a) A hypergraph $H$; (b) the Jaccard weighted line graph $L_J(H)$; (c) the dual hypergraph $H^*$; and (d) the Jaccard weighted clique expansion $Q_J(H)$.}
    \label{fig:hypergraph-simple}
\end{figure}

\noindent Shown in~\autoref{fig:hypergraph-simple}a, is an example hypergraph $H$ with 5 black vertices and 3 colored hyperedges. 
It is specified by $V = \{v_1, \cdots, v_5\}$ and $E = \{e_1, e_2, e_3\} = \{\{v_1, v_2, v_3\}, \{v_2, v_3\}, \{v_3, v_4, v_5\}\}$. 
In most scenarios of this paper, a hypergraph is shown with the hybrid visualization, by superimposing Venn diagram and bipartite graph based visual encodings; black nodes are vertices, colored nodes and colored convex hulls represent hyperedges. 
For instance, in~\autoref{fig:hypergraph-simple}a, the blue hyperedge $e_3$ containing three black vertices ($v_3$, $v_4$, and $v_5$) is represented by a blue node as well as a blue convex hull. 

As part of the pipeline for hypergraph simplification, we convert the hypergraph into a graph. 
There are two candidate graph representations of a hypergraph: the \emph{line graph}~\cite{BermondHeydemannSotteau1977} and the \emph{clique expansion}~\cite{ZienSchlagChan1999}.
The line graph $L(H)$ of $H$ is a graph whose vertex set corresponds to the set of hyperedges of $H$;  two vertices are adjacent in $L(H)$ if their corresponding hyperedges have a nonempty intersection in $H$. 
The clique expansion $Q(H)$ of $H$ constructs a graph from a hypergraph by replacing each hyperedge with a clique among its vertices. 
We formalize these concepts in the following two definitions.

\begin{definition}
The \emph{line graph} $L(H)$ of a hypergraph $H$  consists of a vertex set $\{e^*_1, \cdots, e^*_m\}$, and an edge set $\{(e^*_i, e^*_j) \mid e_i \cap e_j \neq \emptyset, i \neq j\}$. 
\end{definition}

\begin{definition}
The \emph{clique expansion} $Q(H)$ of a hypergraph $H=(V, E)$ consists of vertex set $V$ (the same vertex set as $H$), and there is an edge $(v_i, v_j)$ in $Q(H)$ if there exists some hyperedge $e \in E$ such that $v_i, v_j \in e$.
\end{definition}

The line graph and clique expansion are related through the concept of duality.
\begin{definition}
The \emph{dual hypergraph} $H^*=(E^*, V^*)$ of $H = (V, E)$ has vertex set $E^* = \{e^*_1, \cdots, e^*_m\}$ and hyperedge set $V^* = \{v^*_1, \cdots, v^*_n\}$, where $v^*_i = \{e^*_j \mid v_i \in e_j \text{ in } H\}$. 
\end{definition} 
As shown in~\autoref{fig:hypergraph-simple}c, $H^*$ swaps the roles of vertices and hyperedges. 
For instance, hyperedge $e_1$ in $H$ becomes vertex $e_1^*$ in $H^*$, and vertex $v_1$ in $H$ becomes hyperedge $v_1^*$ in $H^*$.
It is not difficult to show that the clique expansion is the line graph of the dual, $Q(H) = L(H^*)$.

For our hypergraph simplification we work primarily with a weighted line graph or clique complex, using intersection sizes or Jaccard indices as weights.
In $L(H)$ the intersection weight of edge $(e_i^*, e_j^*)$ is $|e_i \cap e_j|$ while the Jaccard weight is $|e_i \cap e_j|/|e_i \cup e_j|$.
By duality, in $Q(H)$ the intersection weight of edge $(v_i, v_j)$ is $|v_i^* \cap v_j^*|$ and the Jaccard weight is $|v_i^* \cap v_j^*|/|v_i^* \cup v_j^*|$. 

The Jaccard weighted line graph, $L_J(H)$, of $H$ is shown in~\autoref{fig:hypergraph-simple}b. 
The hyperedges $e_1$ and $e_2$ in $H$ turn into vertices $e^*_1$ and $e^*_2$ in $L_J(H)$ with weight on $(e^*_1, e^*_2)$ equal to $|e_1 \cap e_2|/|e_1 \cup e_2| = |\{v_2,v_3\}|/|\{v_1, v_2, v_3\}| = 2/3$. 

The Jaccard weighted clique expansion $Q_J(H)$ is shown in ~\autoref{fig:hypergraph-simple}d. 
For example, vertices  $v_1$ and $v_2$ in $H$ belong to sets of edges $\{e_1\}$ and $\{e_1, e_2\}$ respectively. 
The edge  $(v_1, v_2)$ in $Q_J(H)$ has a weight equal to $|\{e_1\} \cap \{e_1, e_2\}|/|\{e_1\} \cup \{e_1, e_2\}| = 1/2$.

In a nutshell, the Jaccard weighted line graph of a hypergraph captures the  similarities between hyperedges; the higher the weights, the more similar they are. 
On the other hand, the Jaccard weighted clique expansion -- the line graph of the dual hypergraph -- captures similarities between vertices. 

\subsection{$s$-Walks and $s$-Connected Components}
\label{subsec:walks} 

For a graph $G = (V,E)$, a \emph{walk} of length $k$ is a sequence of vertices connected by edges. It can also be described as a sequence of successively incident edges. 
We include a similar notion of walk on a hypergraph, introduced by~\cite{aksoy2020hypernetwork}, using a hyperedge perspective.
\begin{definition}
An \emph{$s$-walk of length $k$} between hyperedges $f$ and $g$ in a hypergraph $H$ is a sequence of hyperedges, 
$f = e_{i_0}, e_{i_1}, \cdots, e_{i_k} = g$, where for each $1\leq j \leq k$, $i_{j-1} \neq i_j$ and $|e_{i_{j-1}} \cap e_{i_j}| \geq s$.
\end{definition}

\begin{definition}
For a hypergraph $H = (V, E)$, a subset of hyperedges $C \subseteq E$ is \emph{$s$-connected} if there exists an $s$-walk between all pairs of hyperedges in $C$. 
$C$ is an \emph{$s$-connected component} if it is maximal, that is, there is no $s$-connected set $C' \subseteq E$ such that $C \subsetneq C'$. 
\end{definition}

\begin{figure}[!ht]
    \centering
    \includegraphics[width=0.9\columnwidth]{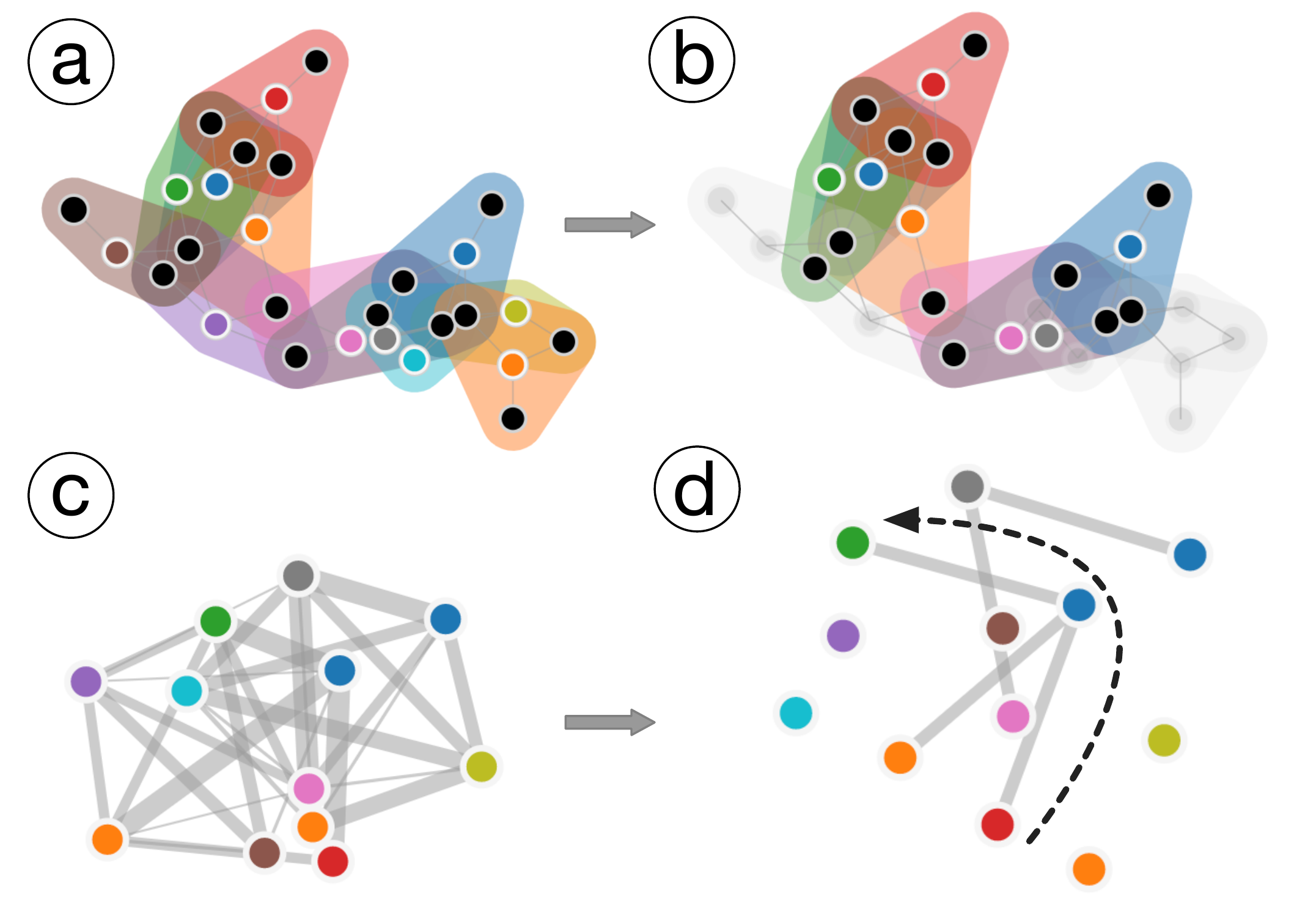}
    \vspace{-2mm}
    \caption{An example of filtering an $s$-line graph using the $s$ parameter, for $s=1$ (c) and $s=3$ (d). Such a filtering leads to a filtering of the original hypergraph (a), where hyperedges that are in singleton $s$-components (i.e., hyperedges that are not $s$-connected to any other hyperedge) are greyed out in (b).}
    \label{fig:multiple-s-walk}
\end{figure}

\begin{definition} 
\label{def:s-line-graph}
An \emph{$s$-line graph} of $H$ (for $s \geq 1$), denoted as $L^s(H)$, is a filtered line graph where edge $(e_i^*, e_j^*)$ is present only if $|e_i \cap e_j| \geq s$. Similarly, an \emph{$s$-clique expansion} of $H$, denoted as $Q^s(H)$, is a filtered clique expansion where edge $(v_i, v_j)$ is present only if $v_i$ and $v_j$ share at least $s$ hyperedges in $H$.
\end{definition}

The notion of $s$-line graphs and $s$-clique expansions allow us to \emph{filter} a hypergraph by its connectivity, as illustrated in~\autoref{fig:multiple-s-walk}. 
For example, there is a $3$-walk (dotted black arrow) between the red hyperdege to the green hyperedge via the blue hyperedge, as shown in~\autoref{fig:multiple-s-walk}d.  

For the remainder of this paper, we work primarily with the Jaccard weighted $s$-line graph of $H$, denoted as $L^s_J(H)$, and the Jaccard weighted $s$-clique expansion of $H$, denoted as $Q^s_J(H)$. Unless otherwise stated, $s=1$. 
In one of our examples we will provide a comparison between edge weights based on the Jaccard indices and the intersection size.

\subsection{Topological Simplifications of Graphs}
\label{sec:bar-graph}
In this section, we first introduce the topological profile of a weighted graph, formally known as its \emph{barcode}~\cite{CarlssonZomorodianCollins2004, Ghrist2008}, which is grounded in persistent homology~\cite{EdelsbrunnerLetscherZomorodian2002}. 
We then use the barcode to guide the topological simplification of the graph. Later, in~\autoref{sec:method-simplify}, we show how simplification of graphs $L^s(H)$ and $Q^s(H)$ are used to perform our hypergraph simplification. 
   
To obtain the barcode of a weighted graph $G$, we apply persistent homology to a metric space representation of the graph~\cite{HajijWangScheidegger2018}. See~\cite{EdelsbrunnerHarer2008} for an introduction and~\cite{CarlssonZomorodianCollins2004} for an algebraic treatment of persistent homology.  
Persistent homology can be used to capture topological features (e.g., connected components, loops, and higher dimensional voids) in any dimension, $d$.
In this paper we will focus on $d=0$, allowing us to simplify the definition of a barcode since this restricted version can be computed using the notion of a minimum spanning tree (MST) of a graph. 
In other words, a merger of two components corresponds to an edge of the MST. 
Recall an MST is a spanning tree with minimum possible total edge weight. 
As an MST can also be used to derive the single linkage clustering (SLC) dendrogram~\cite{GowerRoss1969}, the barcode-guided simplification process is also equivalent to applying the SLC with a threshold.

\begin{figure}[!ht]
    \centering
    \includegraphics[width=0.98\columnwidth]{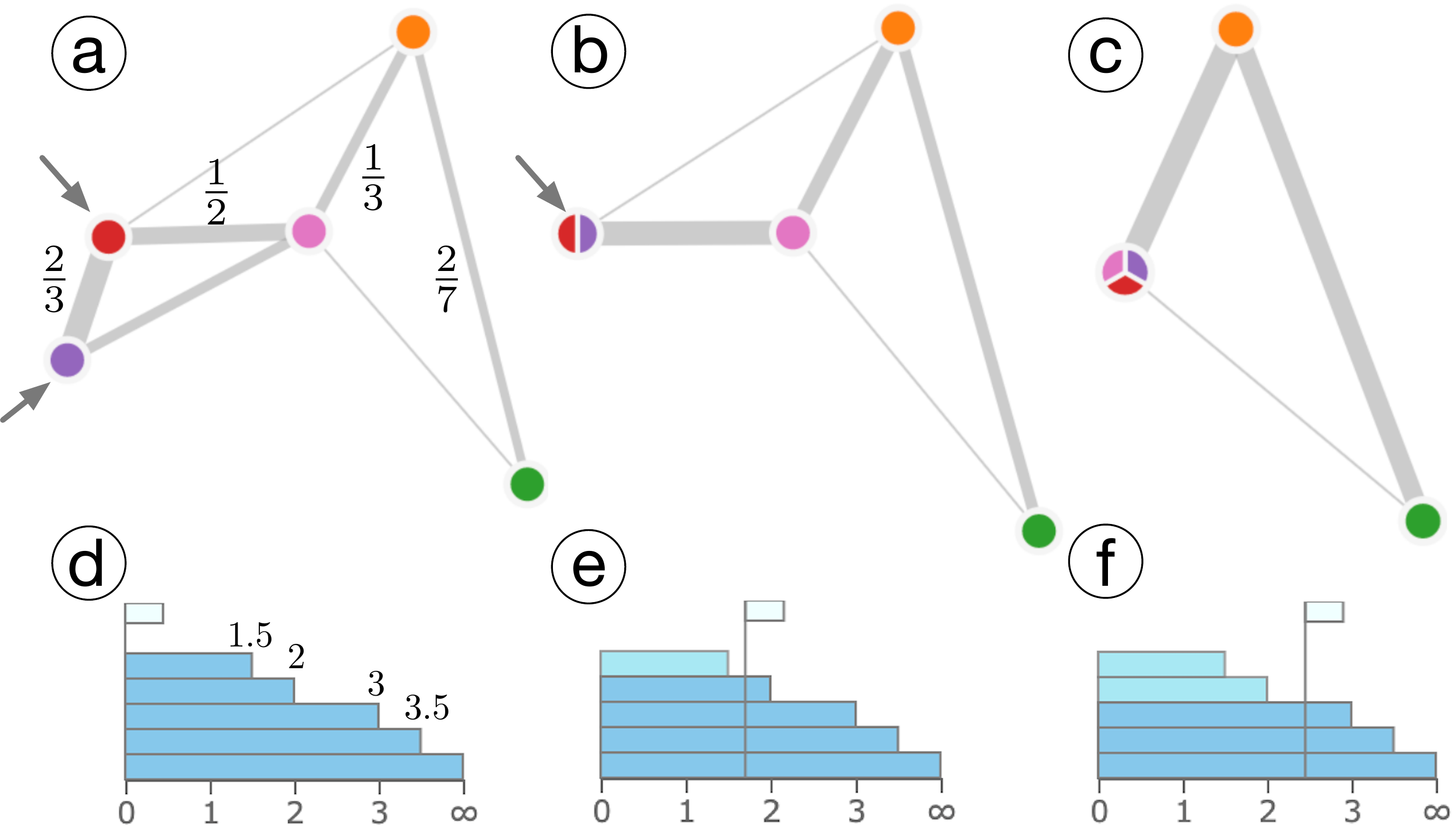}
    \caption{A barcode-guided topological simplification of a graph.}
    \label{fig:graph-simplify}
\end{figure}

A weighted graph, $G=(V, E, w)$, consists of vertices, $V$, edges, $E$, and a weight function $w : E \rightarrow \Rspace^+$. 
Constructing an MST will tend to keep edges with smaller weight and remove those with higher weight. 
Typically this is done because the weights represent distances, where a small weight means two vertices are close, or similar, and a large weight means two vertices are far, or dissimilar.
In the case that weights are similarities, not distances, two vertices with high weight are more similar than two with low weight.
To simplify a similarity-weighted graph, we wish to merge vertices in $G$ into super-vertices based on a decreasing order of their similarities. 
Therefore, before computing the barcode of $G$, we first invert each edge weight $w(e)$ to be $1/w(e)$ and then compute the corresponding MST. 

The \emph{barcode} $\Bcal(G)$ is a visual representation of the MST that consists of a collection of sorted horizontal line segments (\emph{bars}) in a plane, where each line segment (excluding the longest one) corresponds to an edge in the MST with length proportional to its weights. 
As illustrated in~\autoref{fig:graph-simplify}a, four of the thickest edges (shown with edge weights) form the MST; each of these edges gives rise to a bar in the barcode in~\autoref{fig:graph-simplify}d.  
For instance, the edge connecting the most similar vertices -- the red and the purple vertices (pointed by arrows) -- in~\autoref{fig:graph-simplify}a with a weight of $2/3$ gives rise to the shortest bar of length $1.5$ in~\autoref{fig:graph-simplify}d.  

Suh \etal~\cite{SuhHajijWang2019} used the barcode to control the contraction  and repulsion of edges in the force-directed layout of a graph. 
Instead, in this paper, we use the barcode to guide the merging of vertices into super-vertices as part of the simplification pipeline.  
As illustrated in~\autoref{fig:graph-simplify}e, if we choose a threshold that passes the first bar, we combine the purple and red vertices together into a super-vertex in (b). 
Similarly, choosing a threshold that passes the second bar in~\autoref{fig:graph-simplify}f results in the combination of purple, red, and pink vertices into a super-vertex in (c).

\para{Stability.} We consider a barcode-guided simplification of a graph to be mathematically justified in the following sense. 
We call a graph $G'$ an \emph{$\epsilon$-simplification} of another graph $G$, if $G'$ is obtained from $G$ via \emph{vertex contractions} (that is, merging subsets of vertices in $G$ thus contracting the induced edges), and the barcode $\Bcal(G')$ is the same as the barcode $\Bcal(G)$ except all bars with lengths at most $\epsilon$ have been removed. 
Based on the stability of barcodes~\cite{Cohen-SteinerEdelsbrunnerHarer2007}, by performing a simplification up to a threshold of $\epsilon$, the distance between the barcodes of $G'$ and $G$ is upper bounded by $\epsilon$.

Formally, let $\gamma$ be a matching between the intervals (bars) $I$ and $I'$ of  $\Bcal(G)$ and $\Bcal(G')$ respectively, define the bottleneck distance between $\Bcal(G)$ and $\Bcal(G')$ to be 
$$d_\infty(\Bcal(G), \Bcal(G')) = \inf_{\gamma} \sup_{I \in \Bcal(G)} || I - \gamma(I)||_\infty,$$
where $L_\infty$ distance between two bars $I = (b,d)$ and $I ' = (b, d')$ is defined as $|| I - I' ||_\infty = \max(|b-b'|, |d-d'|)$.
In our setting, the $L_\infty$ distance between two bars that start at zero, $I=(0, d)$ and $I'=(0, d')$,  is the absolute difference of their end points, $|| I - I' ||_{\infty} = |d-d'|$.
Then we have the stability of barcodes, 
$$d_\infty(\Bcal(G), \Bcal(G')) \leq \epsilon.$$
By construction, the MST of $G'$ is generated from the MST of $G$ by contracting edges with lengths at most $\epsilon$, therefore merging vertices connected by these edges that are at most $\epsilon$ apart. 
Therefore, $\Bcal(G)$ and $\Bcal(G')$ only differ by the bars that are removed via the simplification, which have lengths at most $\epsilon$.

%% file: sec-methods-simplify.tex
\section{Methods}
\label{sec:method-simplify}

We now describe multi-scale topological simplifications of hypergraphs. 
We use the term vertex (resp.~hyperedge) simplification to mean a sequence of operations that reduce the size of a hypergraph by merging vertices (resp. hyperedges) in decreasing levels of similarity. 
Our framework is as follows:
\begin{itemize}
\item[1.] Map a hypergraph $H$ to a graph representation $G$;
\item[2.] Generate the barcode $\Bcal(G)$ of $G$ and use $\Bcal(G)$ to guide its  simplification;  
\item[3.] A simplified $G$ induces a simplification of $H$. 
\end{itemize}

At the core of our approach is the idea that a simplified clique expansion induces a vertex simplification of the hypergraph, while a simplified line graph induces a hyperedge simplification. 
Using barcodes of these weighted graph representations, we allow vertices to be combined if they belong to almost the same set of edges, and edges to be merged if they share almost the same set of vertices, both in a mathematically justifiable way. 

\begin{figure}[!ht]
    \centering
    \includegraphics[width=0.99\columnwidth]{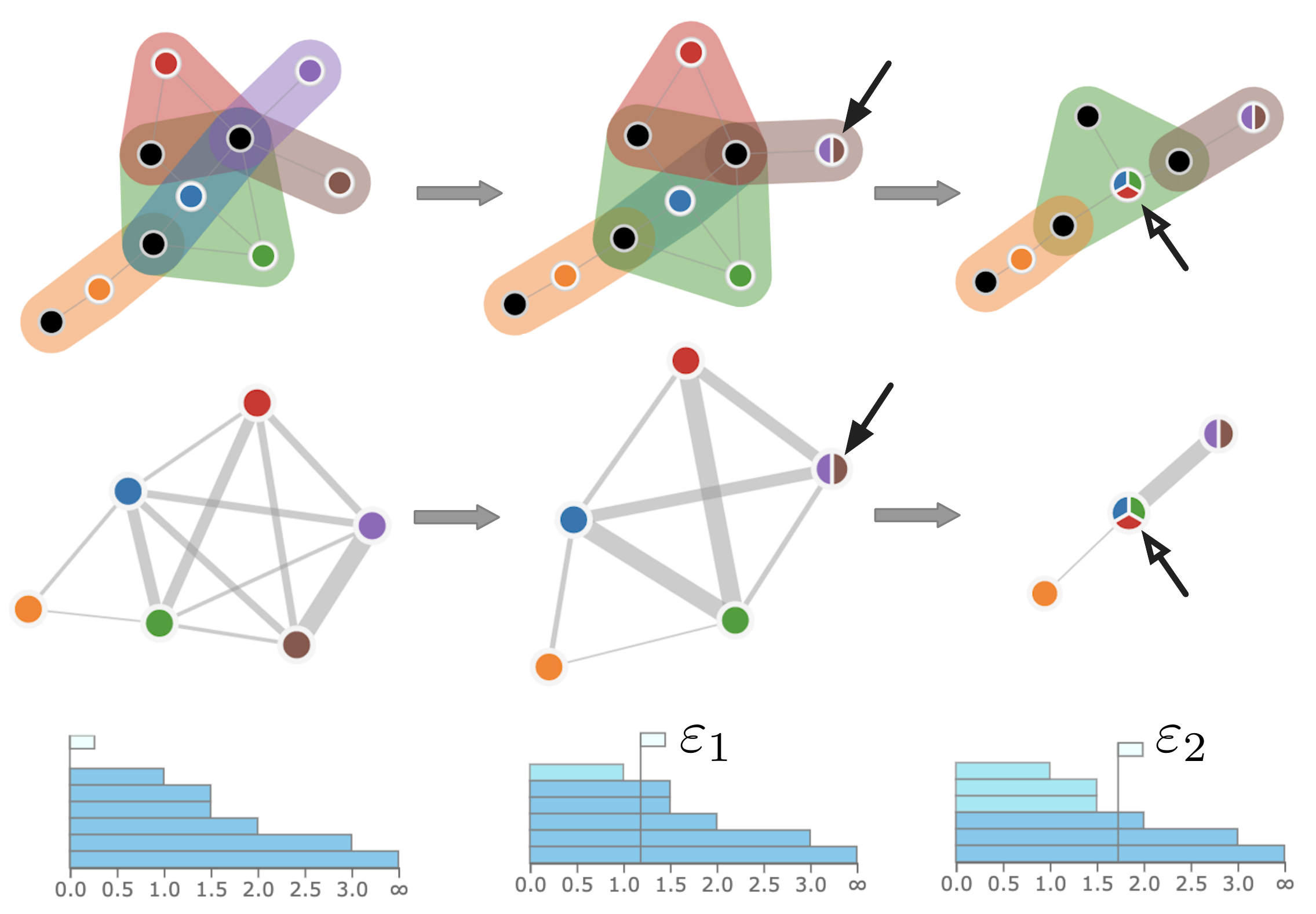}
    \caption{Multi-scale hyperedge simplifications of a hypergraph. Top row includes from left to right: the original hypergraph and its hyperedge simplifications across two scales. Middle row shows its corresponding Jaccard weighted line graphs. Botton row shows the simplification thresholds {\wrt} the barcodes.}
    \label{fig:multi-simplify-edge}
\end{figure}

There are two parameters that guide the simplification process. 
First, the parameter $s$ gives rise to a filtered version of the line graph or the clique expansion. 
~\autoref{fig:multiple-s-walk} illustrates a multi-scale filtering of hyperedges using $s$ parameter, for $s=1$ and $3$ respectively. 

\begin{figure}[!ht]
    \centering
    \includegraphics[width=0.99\columnwidth]{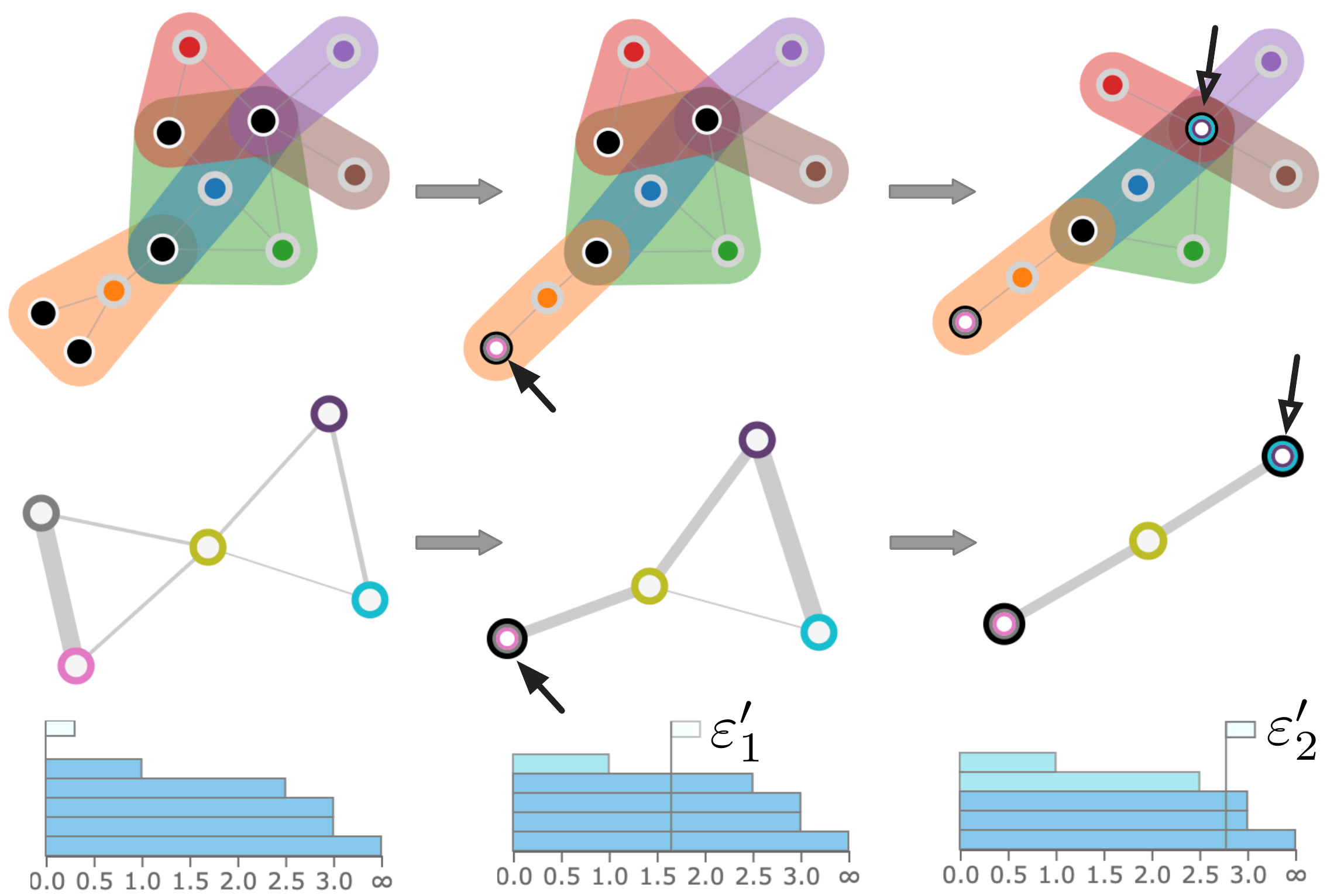}
    \caption{Multi-scale vertex simplifications of a hypergraph. Top row includes from left to right: the original hypergraph and its vertex simplifications across two scales. Middle row shows its corresponding Jaccard weighted clique expansions.}
    \label{fig:multi-simplify-vertex}
\end{figure}

Second, the parameter $\varepsilon$ used in the $\varepsilon$-simplification of a hypergraph $H$ by merging vertices (or hyperedges) whose distances are upper bounded by $\epsilon$ (corresponding, whose similarities are lower bounded by $1/\varepsilon$). 
~\autoref{fig:multi-simplify-edge} illustrates multi-scale hyperedge simplifications.  At $\varepsilon_1$, the brown and purple hyperedges merge into one  (pointed by a black arrow); and at $\varepsilon_2$, three hyperedges in red, green, and blue merge into one at the same time (pointed by the hollow arrow). 
\autoref{fig:multi-simplify-vertex} illustrates multi-scale vertex simplifications. At $\varepsilon'_1$, the grey and pink vertices in the clique expansion (that correspond to the two vertices in the orange hyperedge) merge into one super-vertex (pointed by a black arrow); and at $\varepsilon_2$, two vertices in purple and teal in the clique expansion merge into a super-vertex (pointed by a hollow arrow).

%% file: sec-vis.tex
\section{Interactive Visualization System}
\label{sec:vis}

We provide an open source, interactive visualization system that supports both vertex and hyperedge simplification of an input hypergraph. 
Its user interface is shown in~\autoref{fig:interface}, see the supplementary video for a demo. 
We describe the interface based on hyperedge simplification; the interface for vertex simplification is similar with minor modifications. 

In the middle, the \textbf{graph visualization panel} visualizes the original hypergraph (a), the weighted line graph representation (b), the simplified line graph (d), and the induced simplified hypergraph (c), in \autoref{fig:interface} respectively.  
The vertices and (hyper)edges across (a-d) are connected via linked views based upon their correspondences. 
When we switch from hyperedge simplification to vertex simplification, weighted line graph representations (b, d) become weighted clique expansions accordingly. 

\begin{figure}[!ht]
\centering
\includegraphics[width=1.0\columnwidth]{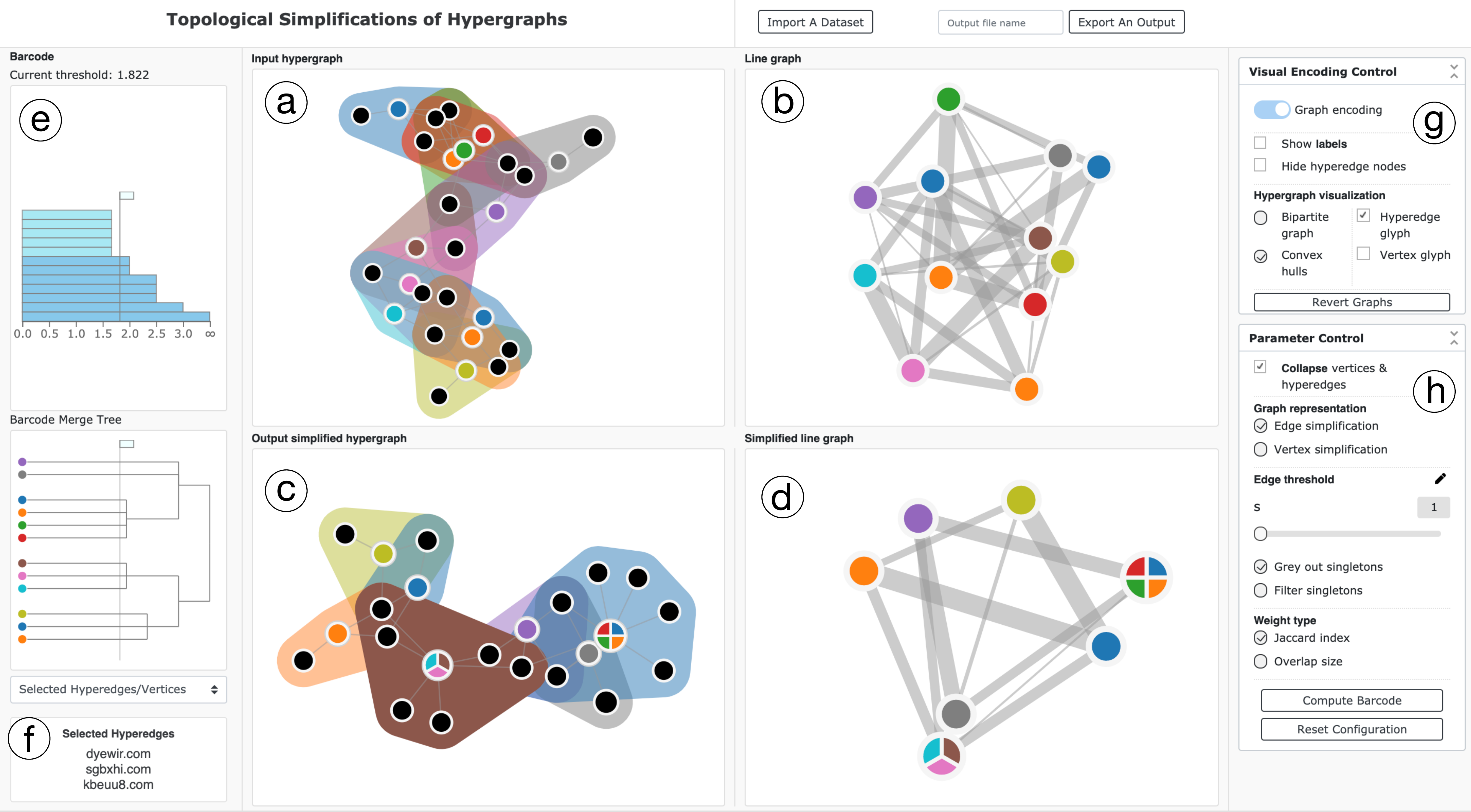} 
\vspace{-6mm}
\caption{Interactive user interface.}
\label{fig:interface}
\end{figure}

On the left, the \textbf{barcode panel} (e) controls the level of simplification (parameter $\epsilon$) for the line graph where vertices connected by edges $\{e^*_i, e^*_j\}$ with weight below the selected threshold in (b) are combined into super-vertices in (d). 
Correspondingly, hyperedges $e_i, e_j$ in (c) are merged into one hyperedge. 
Panel (e) also shows the hierarchical merging of hyperedges (or vertices) via a dendrogram. 
It further performs refinement of a simplified hypergraph via the \emph{bar expansion}: clicking on a bar that has already been simplified will undo the simplification step. 
This means, if hyperedges $e_1$ and $e_2$ have been merged into a single hyperedge $e$ under the current simplification level, clicking the bar that corresponds to this operation will separate $e$ back into $e_1$ and $e_2$. 
To choose the right simplification level, the lower part (f) of the panel displays the labels of hyperedges (or vertices) when hovering on a simplified hyperedge (or vertex). 
Alternatively, it can also display the persistence graph (not shown here) that shows the number of features (connected components) as a function of the persistence level $\epsilon$. 
An appropriate value of $\epsilon$ is typipcally obtained at the plateau of the persistence graph (see \cite{GerberBremerPascucci2010} for details). 

On the right, the \textbf{visual encoding control panel} (g) provides various visual encodings for the hypergraph, see~\autoref{fig:visual-encoding} for an example.
The panel also provides options to display vertex and hyperedge labels. 
The \textbf{parameter control panel} (h) deals with parameter configuration. 
A large input hypergraph can be pre-processed to allow vertex collapse and hyperedge collapse. 
It is important to point out that collapsing the hypergraph here affects the weights in the line graph or clique expansion, which in turn affects the barcode. If two edges intersect in $k$ vertices, and those $k$ vertices are collapsed into $\ell < k$ super-vertices, then the weight between those hyperedges will be $\ell$, not $k$.
Choosing edge simplification employs the line graph representation, while vertex simplification uses the clique expansion.
The $s$ parameter controls the edges present in the $s$-line graph or $s$-clique expansion. 
Singletons in the $s$-line graph or $s$-clique expansion are either greyed out (used in barcode computation but visualized as light gray) or filtered (removed from the hypergraph visualization and not used in barcode computation). 
In computing the barcodes, either Jaccard index or overlap size can be used. 
All of the parameters set in panel (h) contribute to barcode computation. Any change made to the parameters requires a re-computation of the barcode by clicking \emph{compute barcode}.

\para{Implementation.}
We implemented our framework as an interactive web application with a \textit{Python} back-end using a \textit{Flask}-based server. We use the standard \textit{HTML/CSS/Javascript} stack in tandem with \textit{D3.js} and \textit{JQuery} \textit{JavaScript} libraries for designing the user interface and visualization panels. 
The front-end handles data upload, graph and hypergraph visualization, and it sends information about parameter modifications to the back-end. 
The \textit{Python} back-end handles data parsing, creating the hypergraph data structures using the \textit{HyperNetX} \cite{HyperNetX} library (including its edge and vertex collapse functions), constructing bipartite graphs, computing the persistence barcodes, and parameter updates. 
The front-end and back-end communicate using \textit{AJAX} requests and the data is transferred using \textit{JSON} format.

%% file: sec-results.tex
\section{Results}
\label{sec:results}

We provide five example use cases to show how our barcode-guided hypergraph simplification provides an interpretation of the underlying data and helps with insight discovery. 

\subsection{Southern Women}
\label{sec:southern-women}
\begin{figure}[h]
    \centering
    \includegraphics[width=0.49\textwidth]{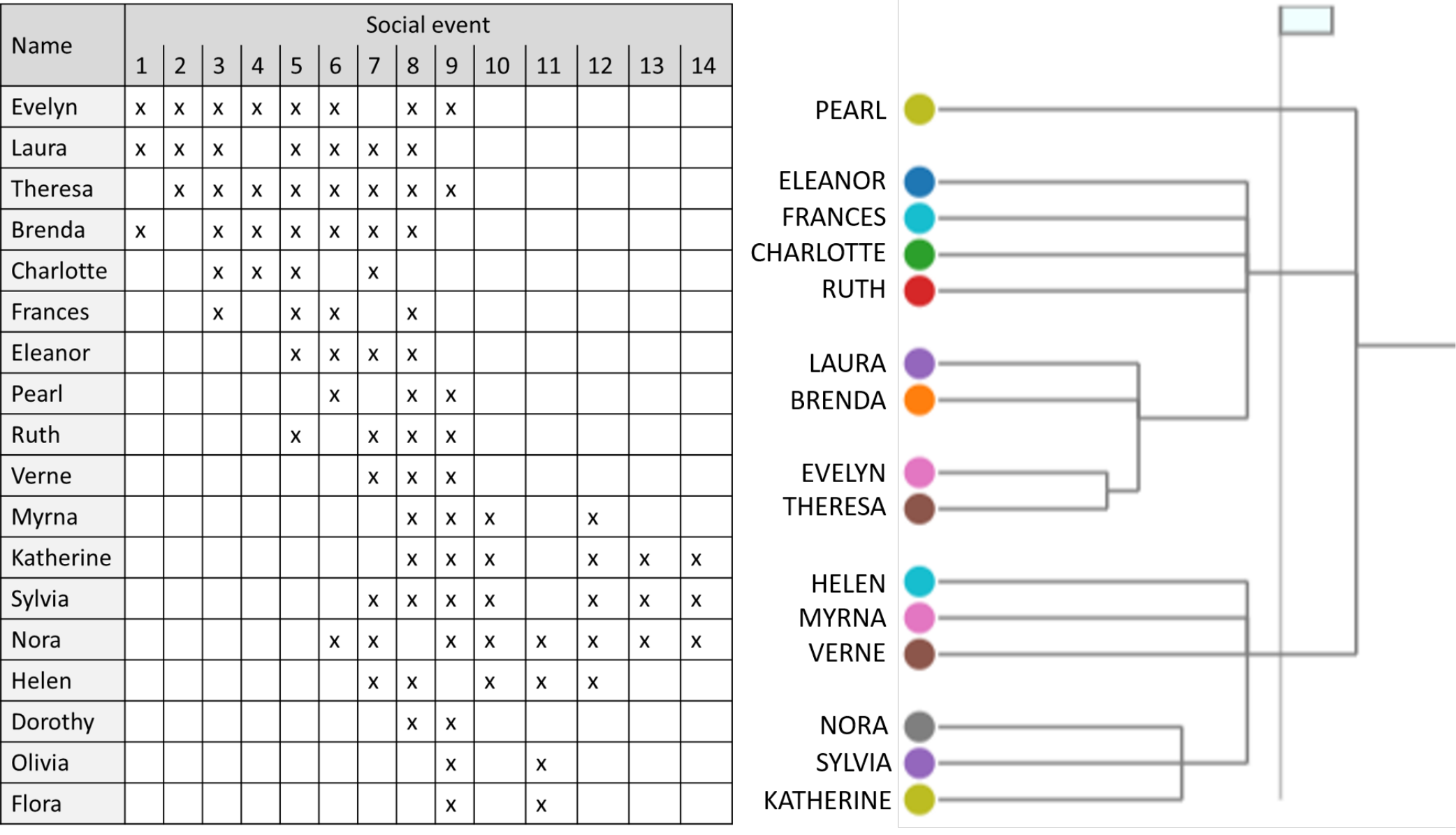}
     \vspace{-4mm}
    \caption{Left: a table reproduced from~\cite{DavisGardnerGardner2009} that records which social events (columns) each woman (rows) attended.
     Right: a hierarchical representation of the simplification (at $s=1$, $\epsilon=0.28$) that highlights intrinsic group structure.}
    \label{fig:sw-table}   
\end{figure}

Our first example considers a small social network.   
In the 1930s, a group of ethnographers collected data on a group of 18 women in Natchez, Mississippi~\cite{DavisGardnerGardner2009}. 
They recorded attendance at 14 informal social events over the course of a nine month period; see~\autoref{fig:sw-table}. 
This dataset has been studied by many other researchers in sociology, information theory, and mathematics, see~\cite{Freeman2003} for a meta-analysis of previous studies. 

\begin{figure}[h]
    \centering
     \vspace{-4mm}
    \includegraphics[width=0.49\textwidth]{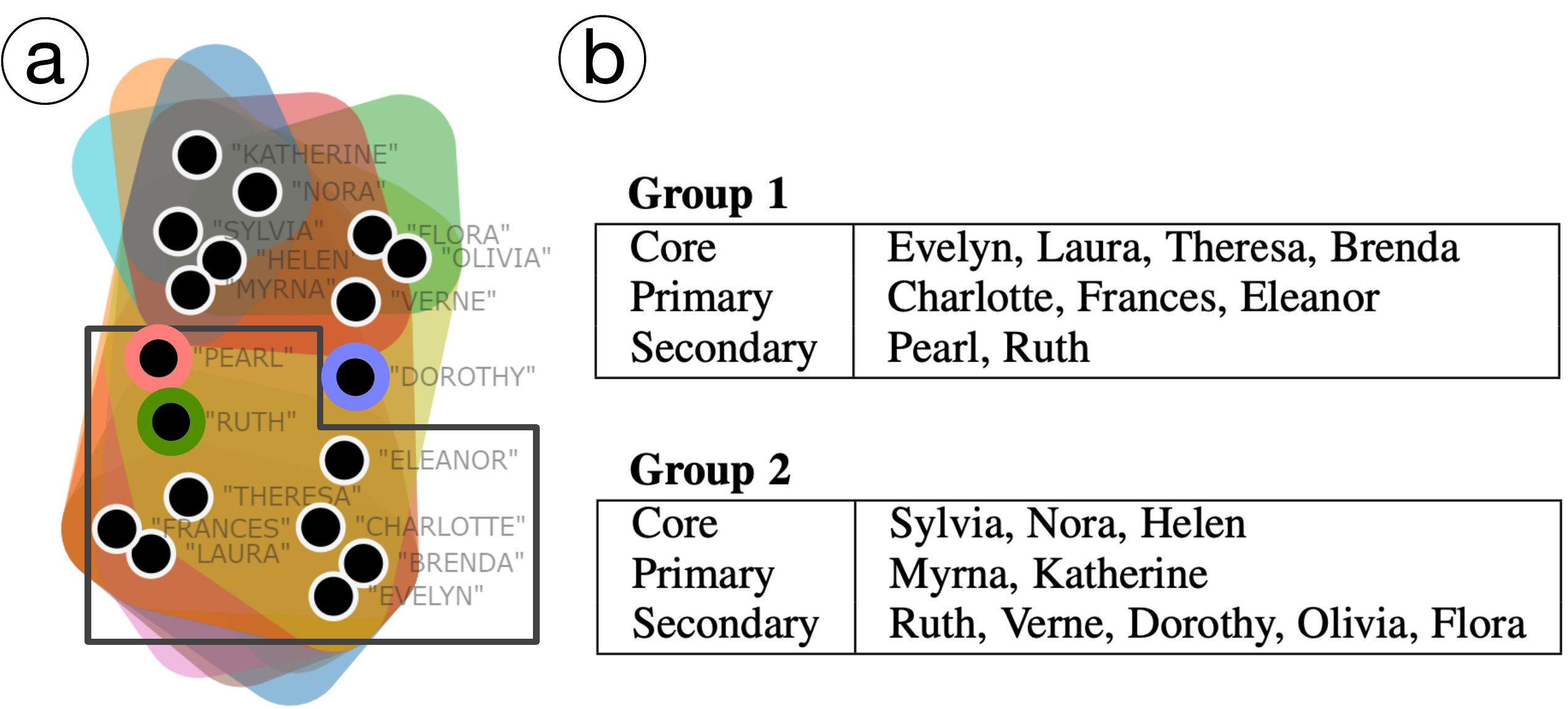}
    \vspace{-6mm}
    \caption{(a) A hypergraph showing women as vertices (black nodes) grouped by events they attended as hyperedges (colored convex hulls). (b) Groups identified from interviews in the original study.}
    \label{fig:sw-hypergraph}
\end{figure}
 
The hypergraph of this dataset is shown in~\autoref{fig:sw-hypergraph}a.
Through interviews with these women, Davis \etal~\cite{DavisGardnerGardner2009} identified two largely distinct groups and determined core, primary, and secondary members of each group based on how they were involved with the events, as summarized in~\autoref{fig:sw-hypergraph}b.
Such groupings are considered as the \emph{ground truth} in our exploration. 
The original layout of the hypergraph in~\autoref{fig:sw-hypergraph}a gives the illusion of a left-right split of the groups, while the ground truth indicates a top-bottom split, where \textbf{Pearl} (red circle) belongs to Group 1, \textbf{Dorothy} (blue circle) belongs to Group 2, and \textbf{Ruth} (green circle) is identified as secondary in both groups.

Using this \emph{Southern Women} dataset, we will demonstrate how  varying parameter choices using our system affects the simplification results, and in some scenarios, highlights the group relations within the ground truth. 
We will observe which parameter choices agree with the reported ground truth and which do not.
In all cases we will perform vertex simplification as the goal is to see how the women are grouped based on the events they attend.

\para{Simplification using Jaccard weights.} 
We first consider vertex simplification of the Jaccard weighted clique expansion $Q^1_J(H)$ derived from the original hypergraph $H$ with $s=1$.

\begin{figure}
    \centering
    \includegraphics[width=0.4\textwidth]{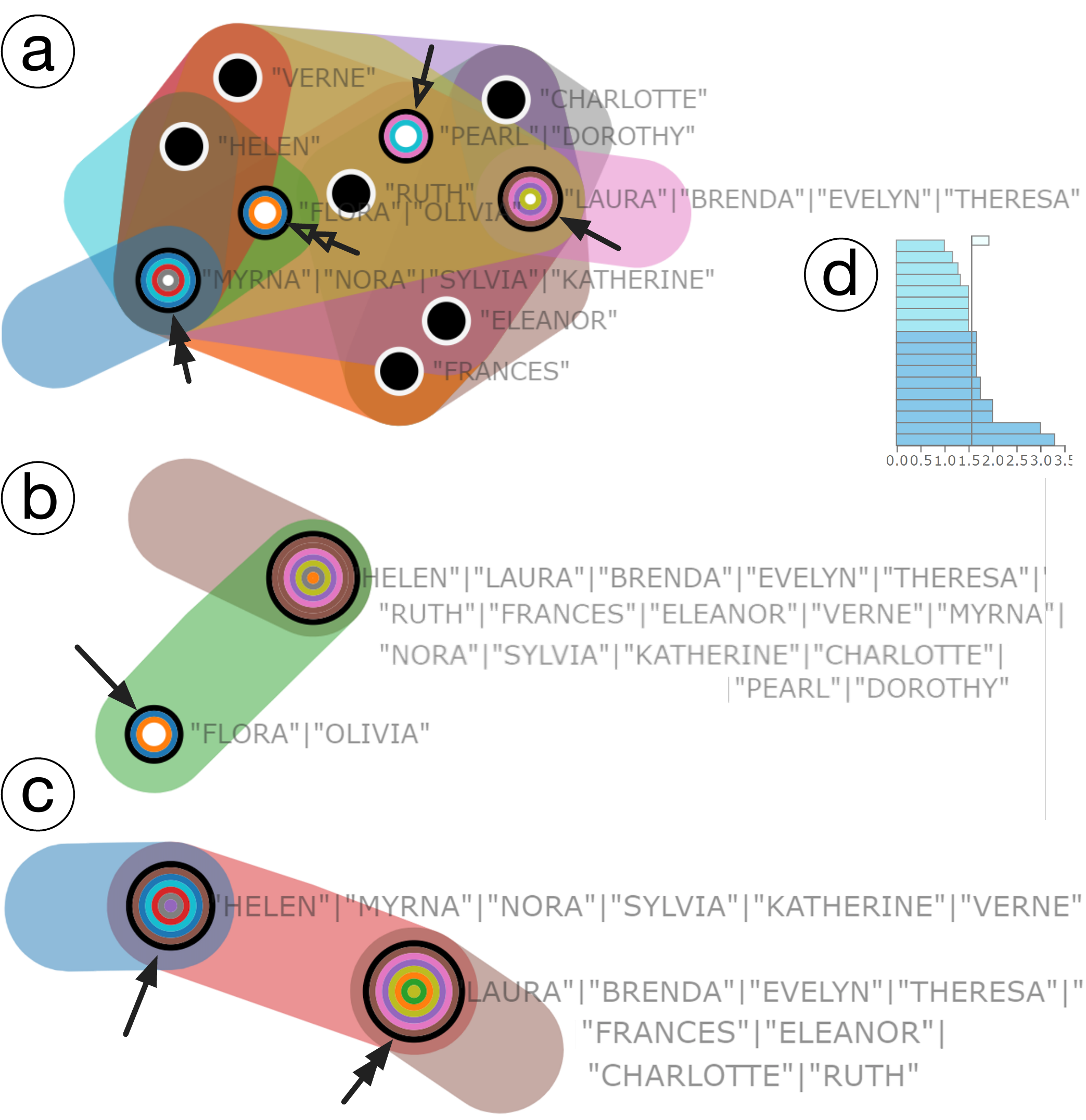}
    \caption{
    Simplification using the Jaccard weighted clique expansion. (a) $s=1$, $\epsilon=1.6$. (b) $s=1$, $\epsilon=3.2$, down to two super-vertices. (c) $s=4$, filtering out singletons, down to two super-vertices. (d) Corresponding barcode for (a).}
    \label{fig:sw-multi-level}
\end{figure}

As we increase our simplification parameter $\epsilon$, guided by \emph{a priori} knowledge of the group structure, we observe that at $\epsilon=1.6$ (\autoref{fig:sw-multi-level}a \& d), super-vertices are formed mostly according to the core groups from the ground truth, with Group 1 core  members on the right (the hyperedge pointed by a filled arrow, \textbf{Laura}, \textbf{Brenda},\textbf{Evelyn}, \textbf{Theresa}), most of the Group 2 core and primary  members on the left (the hyperedge pointed by a double filled arrow, \textbf{Myrna}, \textbf{Nora}, \textbf{Sylvia}, \textbf{Katherine}), and \textbf{Ruth} in the middle. 

As illustrated in~\autoref{fig:sw-multi-level}a, all super-vertices consist of members of the same group, with the exception of \textbf{Pearl} and \textbf{Dorothy} (Group 1 and Group 2 secondary respectively, hollow arrow). 
\textbf{Flora} and \textbf{Olivia} form a hyper-vertex, since they attended exactly the same set of two events (cf.~~\autoref{fig:sw-table}, double hollow arrow).  
All of the core members of Group 1 (\textbf{Laura}, \textbf{Brenda}, \textbf{Evelyn}, \textbf{Theresa}) are merged together into a super-vertex (filled arrow). 
Almost all core and primary members of Group 2 (\textbf{Myrna, Nora, Sylvia, Katherine}, minus \textbf{Helen}) form a second super-vertex (filled double arrow).  
 
We further hypothesize that after simplifying $Q^1_J(H)$ down to two super-vertices, each super-vertex would correspond to a group in the ground truth. 
However, this is not the case as shown in~\autoref{fig:sw-multi-level}b. 
In the simplified hypergraph, one super-vertex contains only \textbf{Flora} and \textbf{Olivia}; while the other super-vertex contains everyone else. 

Finally, we increase the $s$ value to $s=4$, filter out singletons, and simplify $Q^4_J(H)$ to two remaining super-vertices.
The result is shown in~\autoref{fig:sw-multi-level}c.
This filtering process leaves out \textbf{Dorothy}, \textbf{Olivia}, \textbf{Flora}, and \textbf{Pearl} (all of whom are secondary members), but otherwise splits the women into the correct two groups.
The left super-vertex (filled arrow) consists of Group 2 core and primary members and \textbf{Verne} (Group 2 secondary). 
The right super-vertex (filled double arrow) similarly consists of Group 1 core and primary members and \textbf{Ruth} (Group 1 \& 2 secondary).

In summary, simplification using the Jaccard weights and $s=1$ is unable to identify the two groups in the ground truth without using the \emph{a priori} knowledge. 
Using $s=4$, we could identify two subgroups appropriately, however certain secondary members are filtered out unintentionally. 
Next we compare with simplification results using overlap weights.

\begin{figure}
    \centering
    \includegraphics[width=0.4\textwidth]{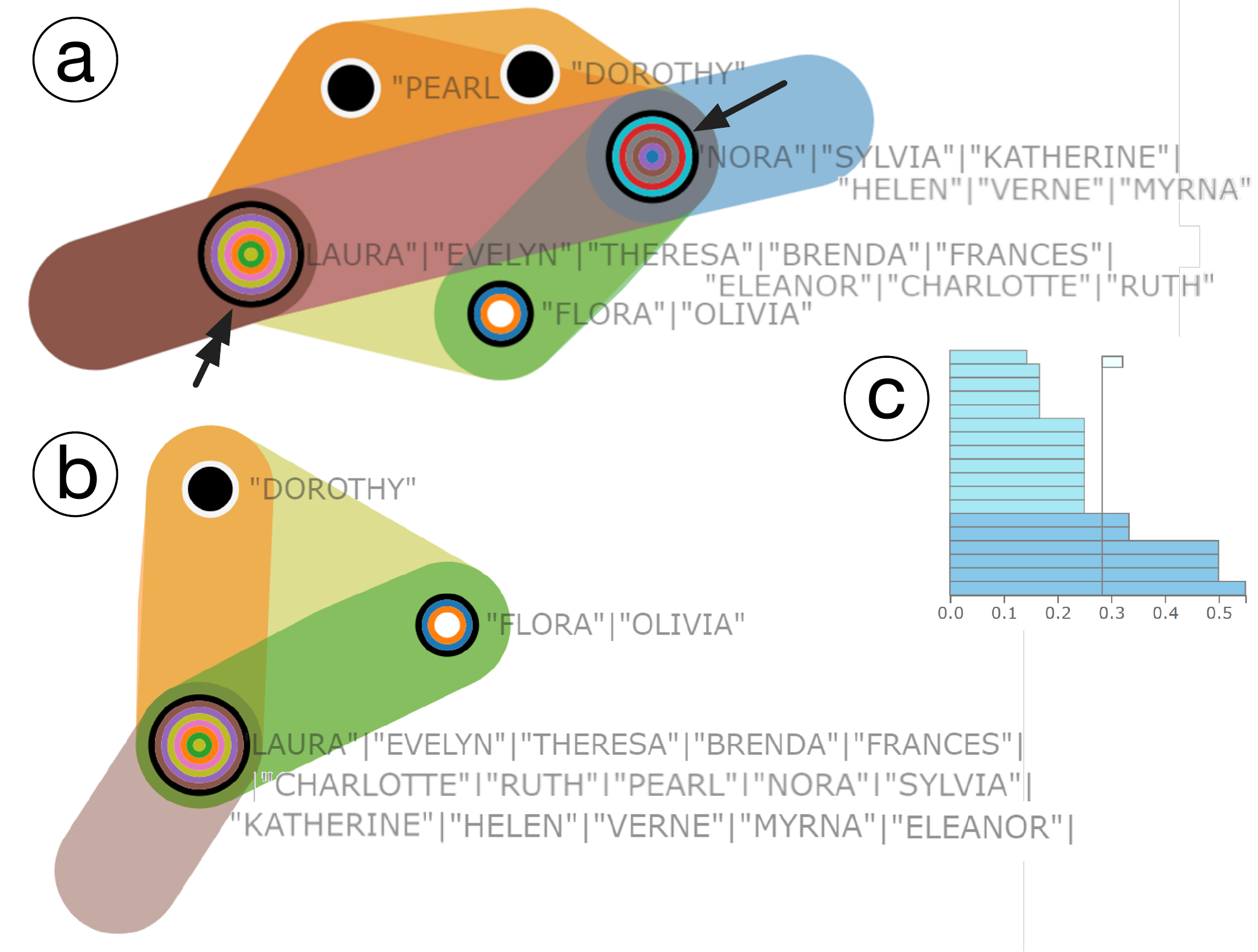}
    \vspace{-2mm}
    \caption{Simplification using overlap weighted clique expansion $Q_w^1(H)$ with $s=1$. (a) $\epsilon=0.28$, with five groups. (b) filtering out singletons, down to three super-vertices.}
    \label{fig:sw-multi-level-overlap}
    \vspace{-4mm}
\end{figure}

\para{Simplification using overlap weights.}  
We first simplify the overlap weighted clique expansion $Q^1_w(H)$ by setting $s=1$. 
As we increase $\epsilon$, the simplification process clearly identifies the two subgroups in the ground truth, see~\autoref{fig:sw-multi-level-overlap}a with the corresponding merging hierarchy in~\autoref{fig:sw-table} (right). 
It identifies the same split as in~\autoref{fig:sw-multi-level}c without filtering \textbf{Pearl, Dorothy, Flora, and Olivia}.  
In particular, the right hyper-vertex (filled arrow) contains all core and primary members of Group 2 plus its secondary member \textbf{Verne}; and the left hyper-vertex (filled double arrows) contains all core and primary members of Group 1 plus \textbf{Ruth}. 
\textbf{Flora} and \textbf{Olivia} are combined as usual.  

As we increase $\epsilon$ further, we simplify $Q^1_w(H)$ down to three super-vertices, see~\autoref{fig:sw-multi-level-overlap}b. 
There is no threshold with two super-vertices as the final simplification merges all three into one super-vertex. 
As in the case of Jaccard weights, this naive simplification does not achieve the desired split into the correct groups.
\textbf{Flora} and \textbf{Olivia} are again grouped together, \textbf{Dorothy} is on her own, and everyone else is grouped together.

Finally, we increase the $s$ value again to $s=4$. 
It shows the same split as \autoref{fig:sw-multi-level-overlap}a, while again  filtering \textbf{Pearl, Dorothy, Flora}, and \textbf{Olivia}.

\para{Final remarks.} 
In this small example with ground truth, we are able to see how Jaccard and overlap weights perform slightly differently. 
The fact that it is easier to identify the two groups using overlap weights may be an artifact of how the two groups were identified in the first place by Davis \etal~through interviews and sociological observations.
Hence we cannot expect such an observation to be generalized to other datasets.
It is clear from this example that Jaccard and overlap weights perform differently and may help provide different insights into the same dataset. 
In crafting the examples in the following subsections we explored each dataset using a variety of parameter and weight choices and will show the choice that provided the most insight.
It happens that Jaccard weights are used in all of the remaining examples.

\subsection{\lesmis}
\label{sec:lesmis}
\begin{figure}[!ht]
\centering
\centering
    \includegraphics[width=0.98\columnwidth]{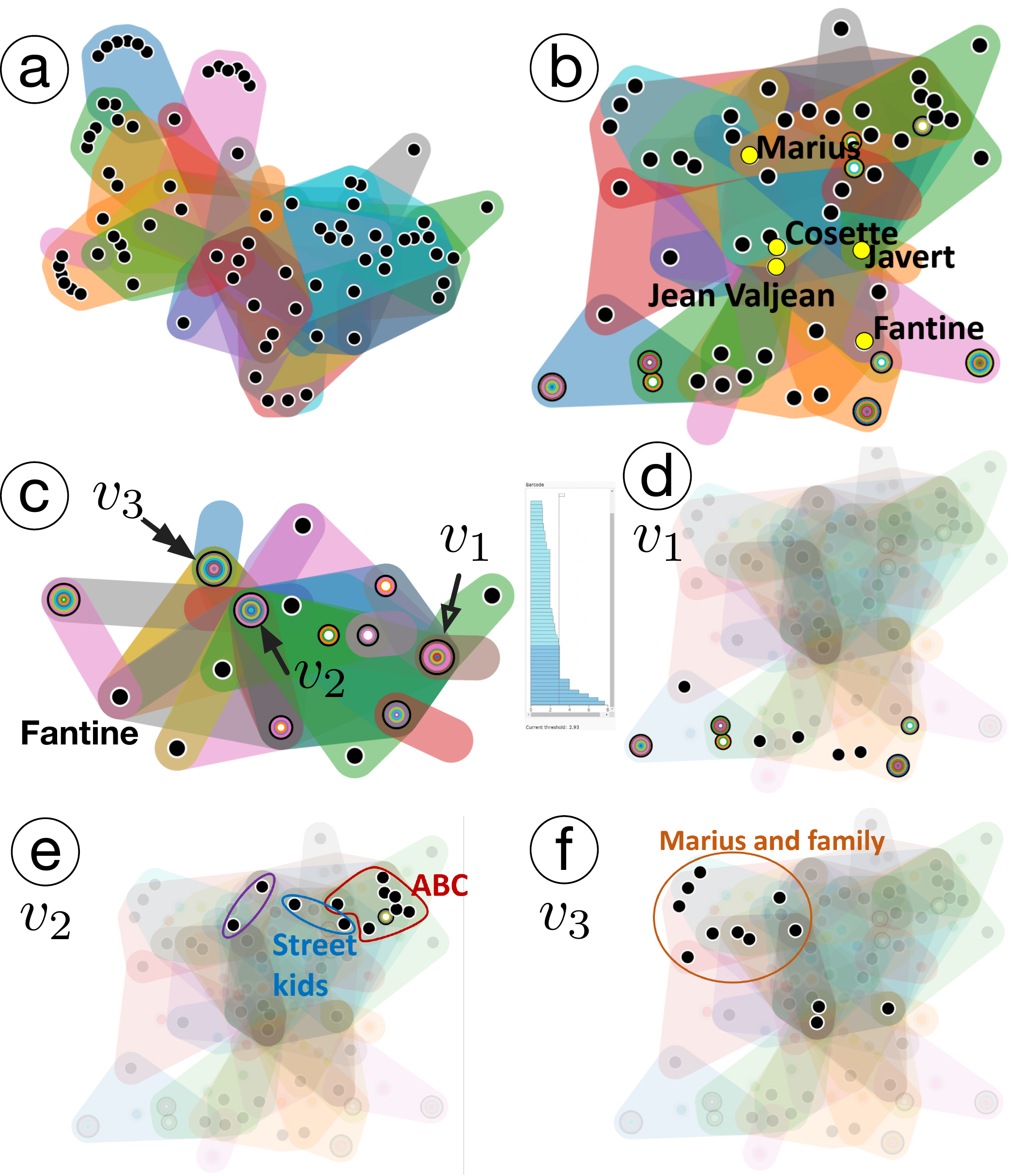}
    \vspace{-2mm}
    \caption{\lesmis{}: hypergraph simplification via vertex simplification, (a) original, 
    (b) collapsed, and (c) simplified hypergraphs with the barcode thresholded at $\epsilon=2.93$. (d-f): Correspondences between three main super-vertices of the simplified hypergraph (c) with their vertices in the original hypergraph (a).}
    \label{fig:lesmis}
\end{figure}

Our second example considers Victor Hugo's novel \emph{\lesmis}, as broken down in the file \texttt{jean.dat} from the Stanford Graph Base~\cite{Knuth1994}. 
This dataset lists the set of characters found within each volume, book, chapter, and scene of the story. 
To form our hypergraph, we consider each character to be a vertex and each (volume, book)-pair to be a hyperedge containing those characters that appear within. 
We use the following parameter settings for the simplification: collapse vertices and edges; vertex simplification; $s=1$; filter singletons; and Jaccard weighted edges. 
\autoref{fig:lesmis} shows the simplification results.

It is not possible to go through the entire plot of this very long novel here, but the story of \lesmis{} revolves around the characters of \textbf{Jean Valjean, Javert, Cosette}, and \textbf{Fantine} (labeled in \autoref{fig:lesmis}b). Other characters of interest include Cosette's love interest \textbf{Marius} (also labeled in \autoref{fig:lesmis}b), a revolutionary student club, and some others that get mixed up in an uprising. 
Our vertex simplification groups many characters together in interesting ways that reflect the narrative of this story. 

In the simplified hypergraph~\autoref{fig:lesmis}c, a vertex (character) of interest is the fact that \textbf{Fantine}, one of the characters many consider a main character in the novel, does not group with her daughter \textbf{Cosette} or any other main character. 
In hindsight, this makes sense, since \textbf{Fantine} appears only in the first volume (of five) and acts as a bridge from the first volume to the rest of the story, losing her daughter \textbf{Cosette} early on. 
\textbf{Fantine} remains an unsimplified vertex (not grouped with any other characters) since she interacts with many groups that do not interact with each other. 
This makes her Jaccard similarity to all of these groups low. 

\autoref{fig:lesmis}(d-f) show and describe correspondences between three of the super-vertices in the simplification ($v_1, v_2$, and $v_3$, pointed by arrows in~\autoref{fig:lesmis}c) with the vertices of the original hypergraph (\autoref{fig:lesmis}a).
The super-vertex $v_1$ from \autoref{fig:lesmis}c contains many peripheral characters in the first volume in \autoref{fig:lesmis}d, those that interact with \textbf{Jean Valjean} and \textbf{Fantine}.
The super-vertex $v_2$ contains all of the ``Friends of the ABC'' revolutionary student group (circled in red) plus two street kids (circled in blue) that get mixed up in the uprising and two additional prominent uprising characters (circled in purple), see \autoref{fig:lesmis}e. 
The super-vertex $v_3 $ from \autoref{fig:lesmis}c contains \textbf{Jean Valjean}, \textbf{Cosette, Javert, Marius}, and all of Marius' family (circled in orange), see \autoref{fig:lesmis}f. 

\begin{figure}[!ht]
\centering
\centering
    \includegraphics[width=0.99\columnwidth]{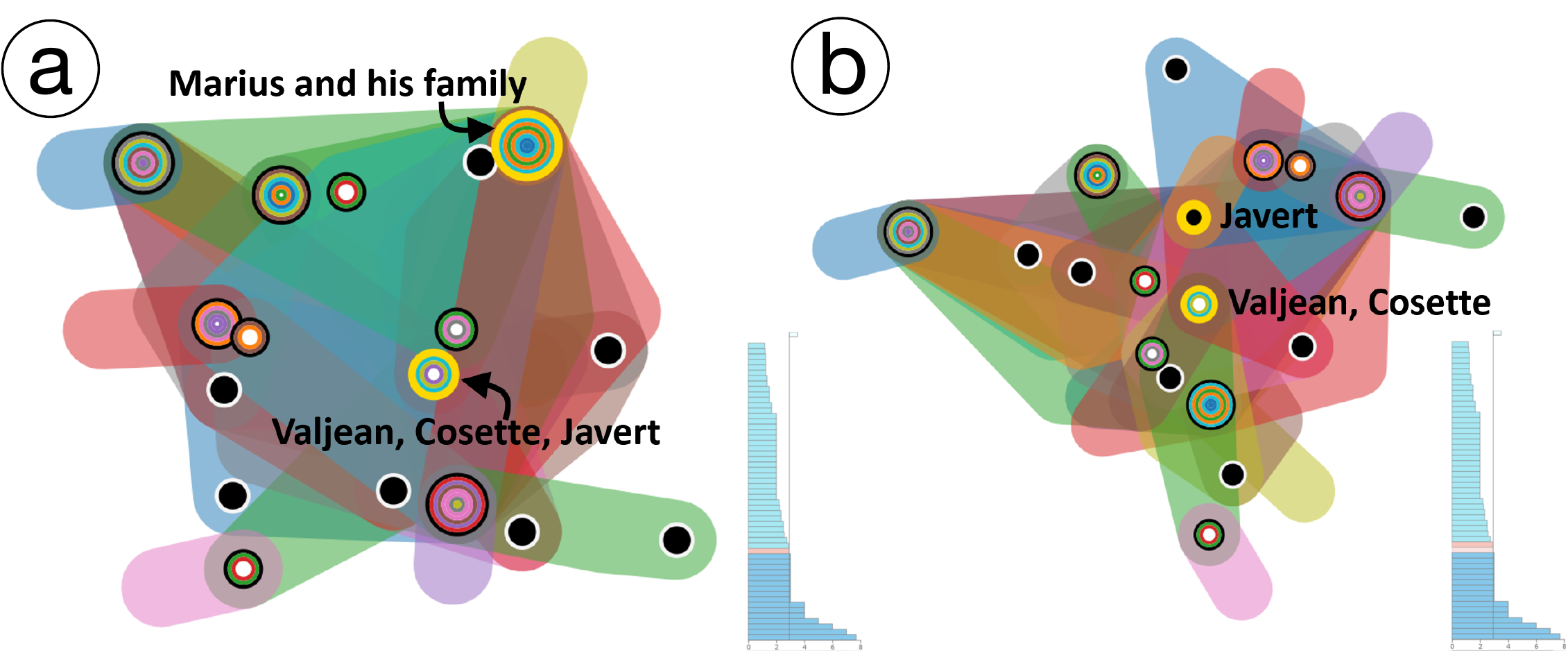}
    \vspace{-2mm}
    \caption{\lesmis{}: expanding the last two bars before the simplification threshold.}
    \label{fig:les-mis-expand}
\end{figure}

Finally, we can use the \emph{bar expansion} capability to explore the two bars that are merged just before our chosen threshold ($\epsilon=2.93$). 
\autoref{fig:les-mis-expand}a shows that expanding one bar splits the super-vertices containing \textbf{Marius} and his family from the one containing \textbf{Jean Valjean, Cosette}, and \textbf{Javert}. 
In \autoref{fig:les-mis-expand}b, we see that expanding the second bar further splits \textbf{Valjean} and \textbf{Cosette} from \textbf{Javert}. 
Both of these operations make sense in the context of the story. 
While all of these characters interact frequently, \textbf{Valjean} and \textbf{Cosette} are certainly the closest and most central pair in the story. 
Moreover, \textbf{Javert} is always chasing \textbf{Valjean} so they do not interact as much, instead they each interact with the same intermediate characters.

While this barcode-guided simplification does not capture the complete narrative flow of the story from beginning to end, it does group characters in ways that align with the story. 
It also shows how various smaller groups interact with each other as smaller groups merge to form larger groups, and how some characters act as bridges between others.

\subsection{Hallmarks Biological Pathways}

Our third example explores the Hallmarks biological pathways hypergraph \cite{LiberzonBirgerThorvaldsdottir2015} as shown in \autoref{fig:s25-e10}. 
The full Hallmarks dataset contains 50 pathways (hyperedges) and 4,386 genes (vertices).
We first use HyperNetX to break this large hypergraph into 2-connected components in order to focus on a smaller subset for visual exploration. We chose a component that contains 10 hyperedges. 
Throughout this example, biological information about pathways is obtained online from MSigDB~\cite{LiberzonBirgerThorvaldsdottir2015,SubramanianTamayoMootha2005}.

\begin{figure}[!ht]
\centering
\centering
    \includegraphics[width=0.99\columnwidth]{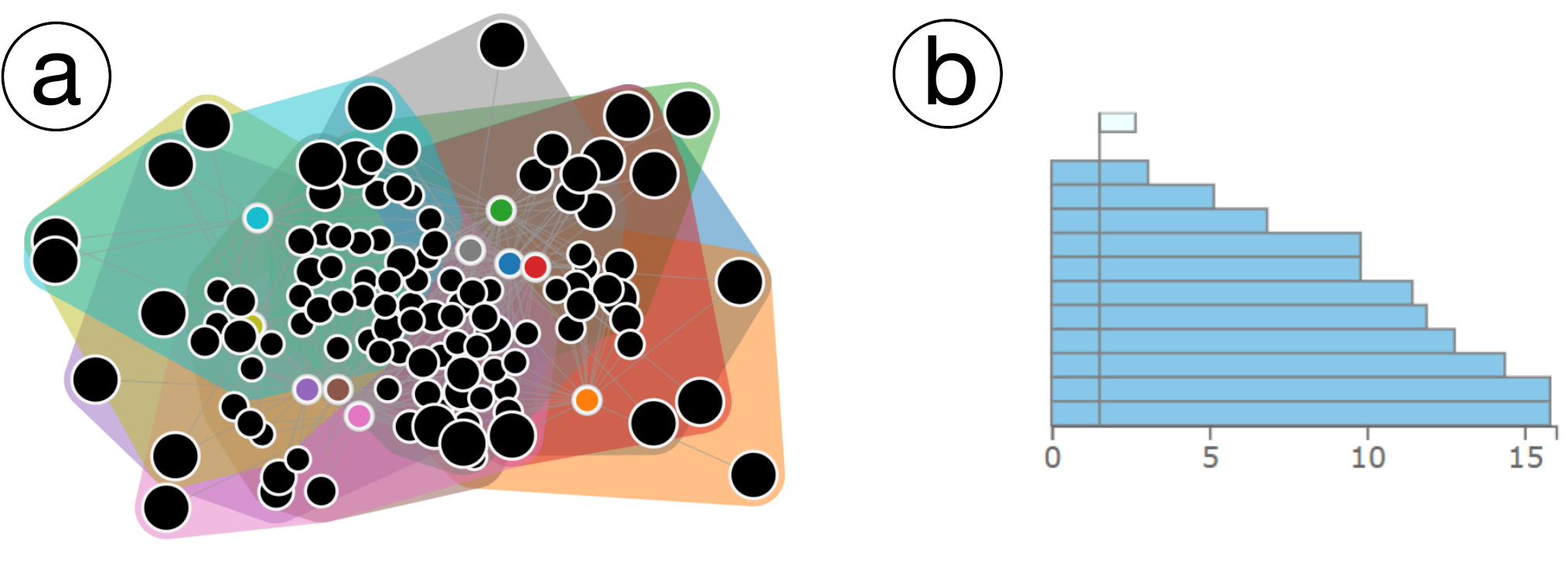}
    \vspace{-2mm}
    \caption{The original Hallmark hypergraph is shown in \autoref{fig:s25-e10}. Here we show (a) the vertex collapsed Hallmark hypergraph, and (b) the barcode for the Hallmark hypergraph.}
    \label{fig:hallmark-start}
\end{figure}

\begin{figure*}[!ht]
\centering
\centering
    \includegraphics[width=0.85\textwidth]{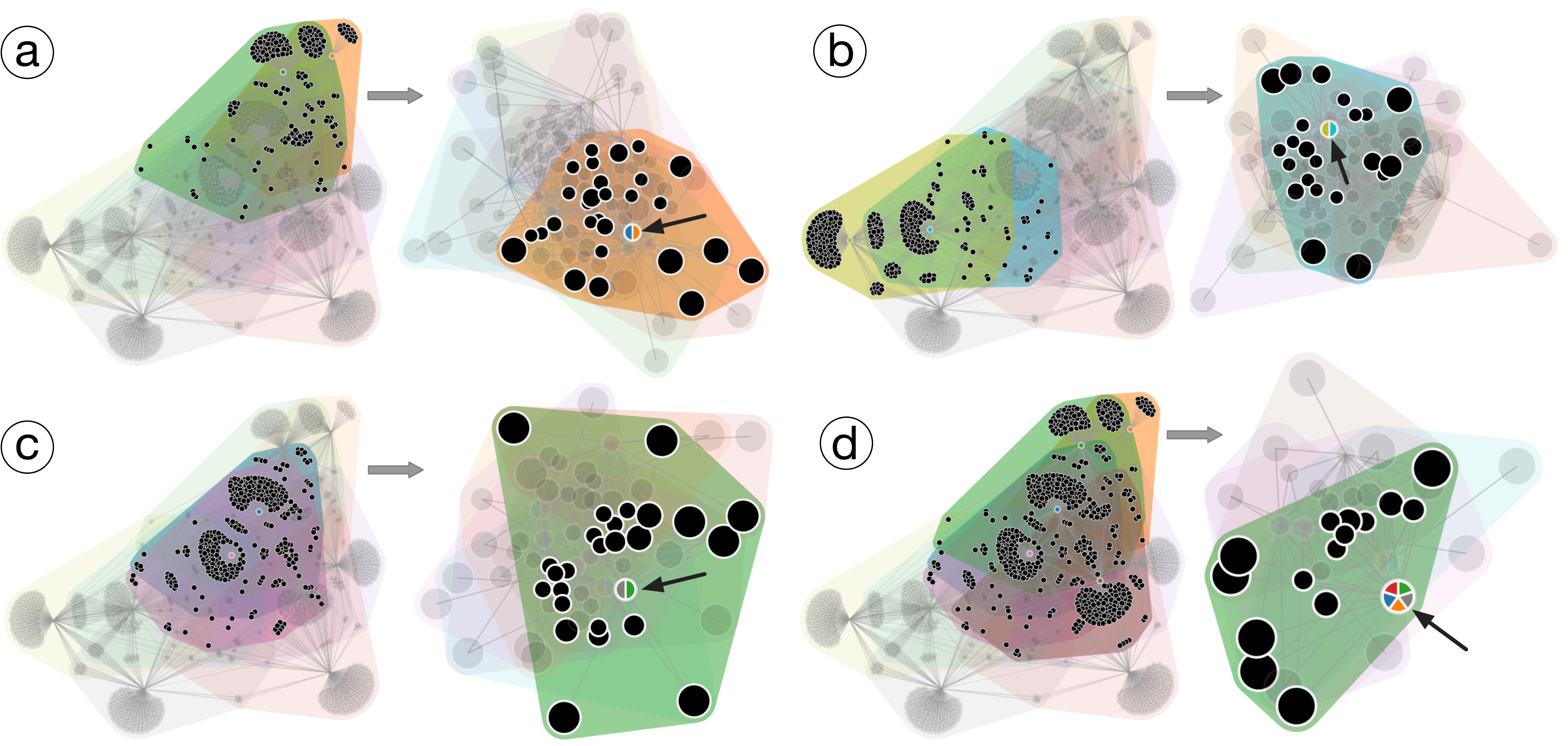}
    \vspace{-2mm}
    \caption{(a-d): Steps 1 to 4 of Hallmark hypergraph simplification. For each subfigure,  right image shows the new merged hyperedge in the simplified hypergraph, left image  highlights the corresponding hyperedges in the original hypergraph.
    (a) Step 1 merges hyperedges (pathways) INTERFERON\_ALPHA and INTERFERON\_GAMMA . 
    (b) Step 2 merges pathways HYPOXIA and GLYCOLYSIS. 
    (c) Step 3 merges pathways INFLAMMATORY\_RESPONSE and TNFA\_SIGNALING. 
    (d) Step 4 merges hyperedges formed in steps 1 and 3 with ALLOGRAFT\_REJECTION.}
    \label{fig:hallmark-process}
\end{figure*}

For this example, we use the following parameter settings: do not collapse vertices or edges; edge simplification; $s=1$; grey out singletons; and Jaccard weights. Rather than choosing one threshold for simplification as in the previous example, we will show insights gathered by walking through the various $p$ thresholds to see the order in which the hyperedges (pathways) merge. 
The original Hallmark hypergraph and the vertex collapsed hypergraph is shown in \autoref{fig:s25-e10}. 
In \autoref{fig:hallmark-start}, we show the vertex collapsed hypergraph by treating collapsed vertices as single vertices along with the initial barcode for this simplification process.

\autoref{fig:hallmark-process}(a-d) shows the first four steps of the simplification process from left to right. 
The 1st two hyperedges (pathways) to merge are \textbf{INTERFERON\_ALPHA\_RESPONSE} and \textbf{INTERFERON\_GAMMA\_RESPONSE} (\autoref{fig:hallmark-process}a). 
Both are pathways that contain genes up-regulated in response to interferon (resp.~alpha or gamma) proteins. 
The 2nd pair (\autoref{fig:hallmark-process}b) to merge are \textbf{HYPOXIA} (genes up-regulated in response to low oxygen) and \textbf{GLYCOLYSIS} (genes involved in breaking down glucose). 
Unlike the 1st pair of edges, the processes of hypoxia and glycolysis are not very related, but they are merged early in the process so they must have many genes in common. 
An observation that may be useful for biology researchers.
The 3rd bar guides the merging of \textbf{INFLAMMATORY\_RESPONSE} with \textbf{TNFA\_SIGNALING\_VIA\_NFKB} (genes regulated by NF-kB in response to TNF), see \autoref{fig:hallmark-process}c. 
The NF-kB protein complex and TNF protein are both known to play a role in regulation of immune response so a merge with an inflammatory response is reasonable. 
The 4th step merges the \textbf{INTERFERON} pair in (a) with the \textbf{INFLAMMATORY} and \textbf{TNFA} pair in (c) and the \textbf{ALLOGRAFT\_REJECTION} pathway, which is involved with transplant rejection. 
This new super-hyperedge seems to encompass all of the inflammatory response edges present within or original set of 10.

At this point, we have five (super-)hyperedges in our simplified hypergraph: inflammation/immune pathways; hypoxia and glycolysis; APOPTOSIS (programmed cell death); MTORC1\_SIGNALING (related to mTORC1 complex involved in protein synthesis);  and P53\_PATHWAY (related to cancer suppression). 
In the next four steps of the simplification (1) APOPTOSIS merges with the inflammation/immune hyperedge, (2) MTORC1 merges with hypoxia and glycolysis, (3) those two just created super-edges merge together, and (4), P53 merges in last.

\begin{figure}
    \centering
    \includegraphics[width=0.35\textwidth]{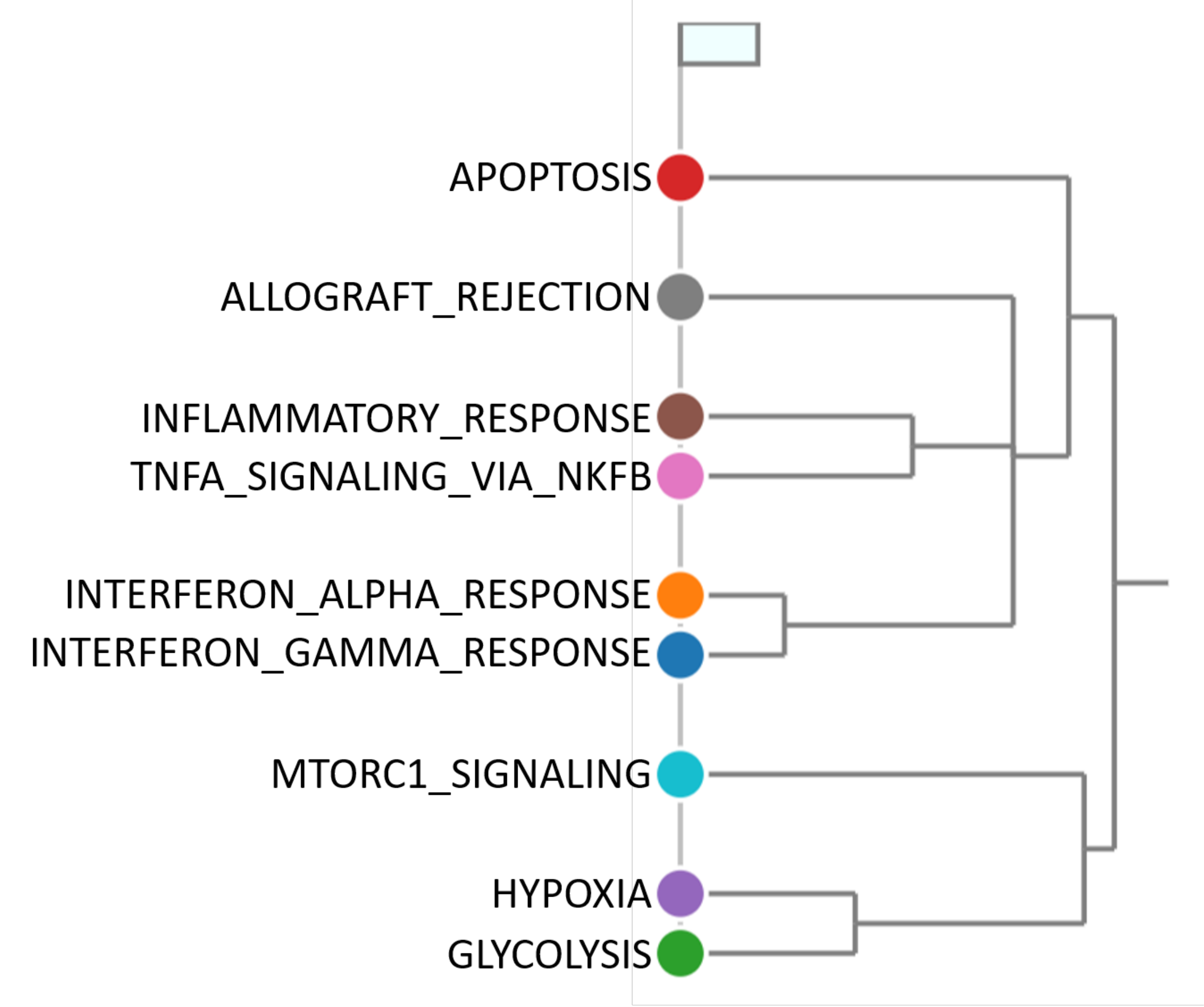}
    \vspace{-2mm}
    \caption{Similarity hierarchy for 10 Hallmark pathways.}
    \label{fig:hallmark-hierarchy}
\end{figure}

This simplification process provides a merging hierarchy amongst these 10 pathways, see \autoref{fig:hallmark-hierarchy}. 
While MSigDB provides pairwise overlap sizes for all of these pathways, the most important insight using our framework is that our simplification process goes beyond pairwise associations and hierarchically merges pathways to discover groups of related pathways. 

\subsection{ActiveDNS}
Our fourth example is from computer networking, specifically the Domain Name System (DNS). 
DNS focuses on how computers translate from the human interpretable domains (e.g., \texttt{www.google.com}) to computer-readable IP addresses (e.g., 192.168.1.1). 
It might seem like these should be one to one, that is, each domain has an IP address, and each IP maps to a domain, but this is not always the case. 
Domain aliasing means that sometimes multiple domains map to the same IP (misspellings like \texttt{www.gogle.com} still get you to the right place) and website hosting services mean that multiple domains can be served up from the same IP address. 
The way that IPs are allocated to domains can show interesting patterns. 
Using a hypergraph representation of the data, where vertices are IP addresses and edges are domains, we ask the question of whether or not domains that share many common IP addresses have any common properties. 

\begin{figure}[!ht]
\centering
\centering
    \includegraphics[width=0.9\columnwidth]{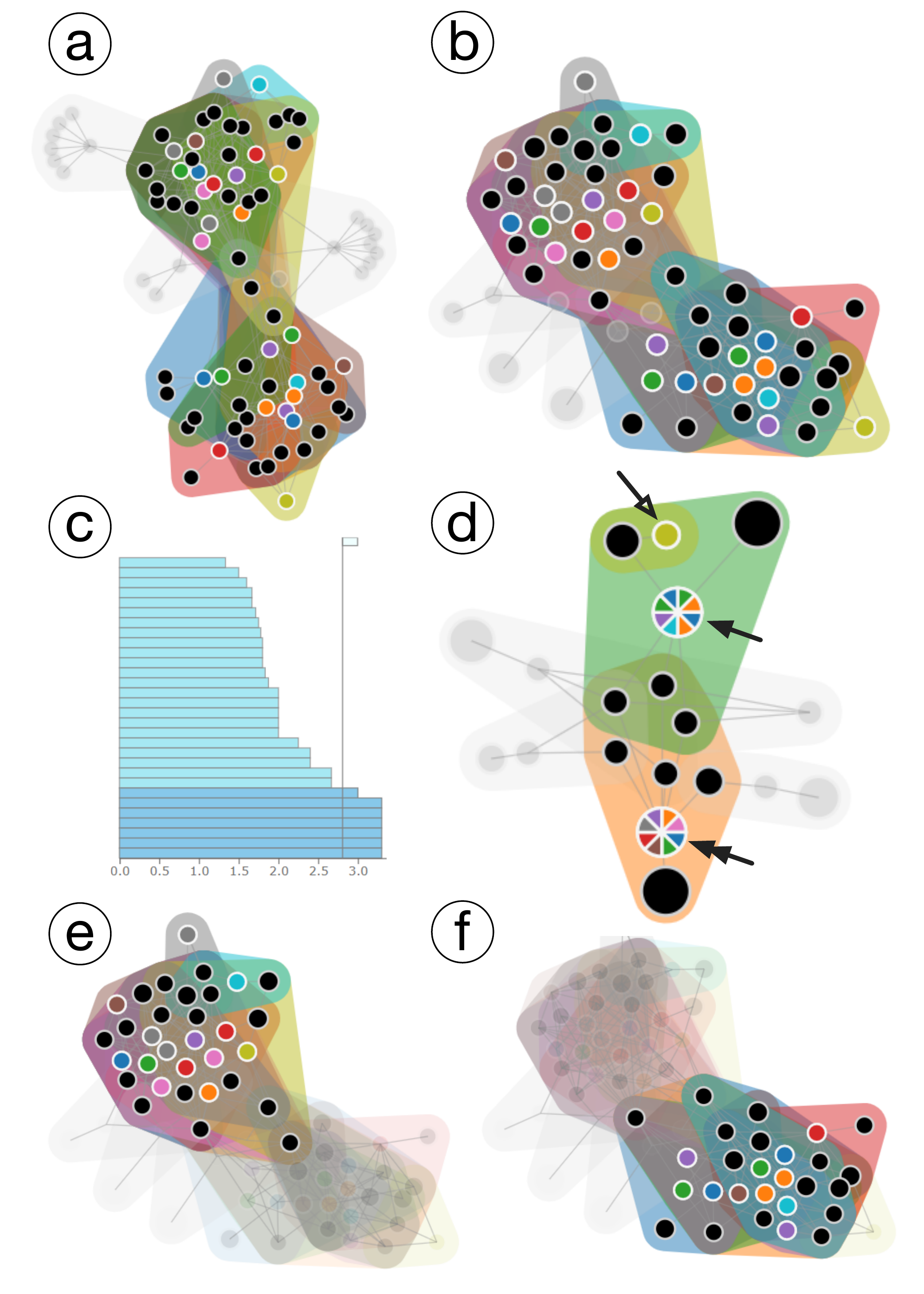}
    \vspace{-2mm}
    \caption{(a-d) ActiveDNS hypergraph simplification. (e-f) Detailed analysis of simplified hyperedges. (e) Hyperedges from (b) that correspond to the green hyperedge in (d). 
    (f) Hyperedges from (b) that correspond to the orange hyperedge in (d).}
    \label{fig:activeDNS}
\end{figure}

The dataset for this exploration comes from ActiveDNS~\cite{kountouras2016enabling}. 
This project out of Georgia Tech does daily active DNS lookups for millions of IP addresses and records the query results in a database. 
We work with one day of DNS records, from April 26, 2018, as our test case. 
This data was also explored as a hypergraph in \cite{JoslynAksoyArendt2020}. 
The entire hypergraph from this day has millions of vertices and edges so we first used the Chapel Hypergraph Library (CHGL)~\cite{CHGL,JeLBhT18} to break the hypergraph into two connected components and then chose one of these components which has 30 hyperedges (domain names) to explore visually using our framework. 
The original hypergraph is shown in \autoref{fig:activeDNS}a and the collapsed in \autoref{fig:activeDNS}b.

For the simplification, we use the following parameters: collapse vertices and edges; edge simplification; $s=3$; grey out singletons; and Jaccard weight. 
We chose $s=3$ to get more to the core of the interactions within this component. 
The barcode (with the chosen threshold), and the resulting simplified hypergraph, are shown in \autoref{fig:activeDNS}c and \autoref{fig:activeDNS}d respectively. 

As shown in \autoref{fig:activeDNS}d, the simplified hypergraph shows a clear separation into two main super-hyperedges, green (filled arrow) and orange (filled double arrow), with one additional yellow hyperedge (hollow arrow) that is fully contained within the green. 
There are also 4 greyed out hyperedges which are not 3-connected to the rest of the component. 
The separation is also evident in the original and collapsed hypergraphs but the groupings become more obvious in the simplified version. 
After making this observation, we use the WHOIS lookup from Hurricane Electric BGP Toolkit~\cite{BGP} to answer the question of what the domains that are grouped have in common. 
WHOIS gives publicly available information about the organization that registered the domain, their contact information, and a variety of other metadata.

\autoref{fig:activeDNS}e highlights the 14 edges in the original hypergraph that are collapsed to form the orange edge in the simplified version. 
The domain names (hyperedges) that are collapsed include \texttt{worldsleadingcruiselines.com}, \texttt{worldsleadingcruiselines.net}, \texttt{wlcl.com}, and \texttt{worldleadingcruiseline.com}. 
In fact, all 14 are some play on ``World's Leading Cruise Lines'' and all domains are registered to Carnival Corporation which is, in fact, one of the world's largest cruise lines. 
Interestingly, one of the grayed out hyperedges has the domain \texttt{comebacktothesea.com}. 
``Come Back to the Sea'' is an older advertising campaign for Carnival and perhaps that is why it does not have as much overlap with the other domains with the more current slogan.

The 11 edges highlighted in \autoref{fig:activeDNS}f, which are collapsed to form the green edge in the simplified version, have less in common on the surface. 
These domains include \texttt{bbgdirect.com}, \texttt{azattykplus.kg}, and \texttt{globalnewsdashboard.com}. 
Within these 11 domains, 7 of them have ``Radio Free Europe'' listed as the registered organization in the publicly available WHOIS information, two have no registered organization, and two have completely different organizations. 
Finally, the yellow edge that is contained within the green in \autoref{fig:activeDNS}c has domain \texttt{radiosvobodakrim.mobi} and is also registered to ``Radio Free Europe.'' 
We make no claims of associations between these domains, especially those registered to other organizations. 
We only observe that our hypergraph simplification method grouped domains together that, for the most part, were registered by the same organization.

\subsection{Coauthor Networks}

\begin{figure}[!ht]
\centering
\vspace{-2mm}
    \includegraphics[width=0.99\columnwidth]{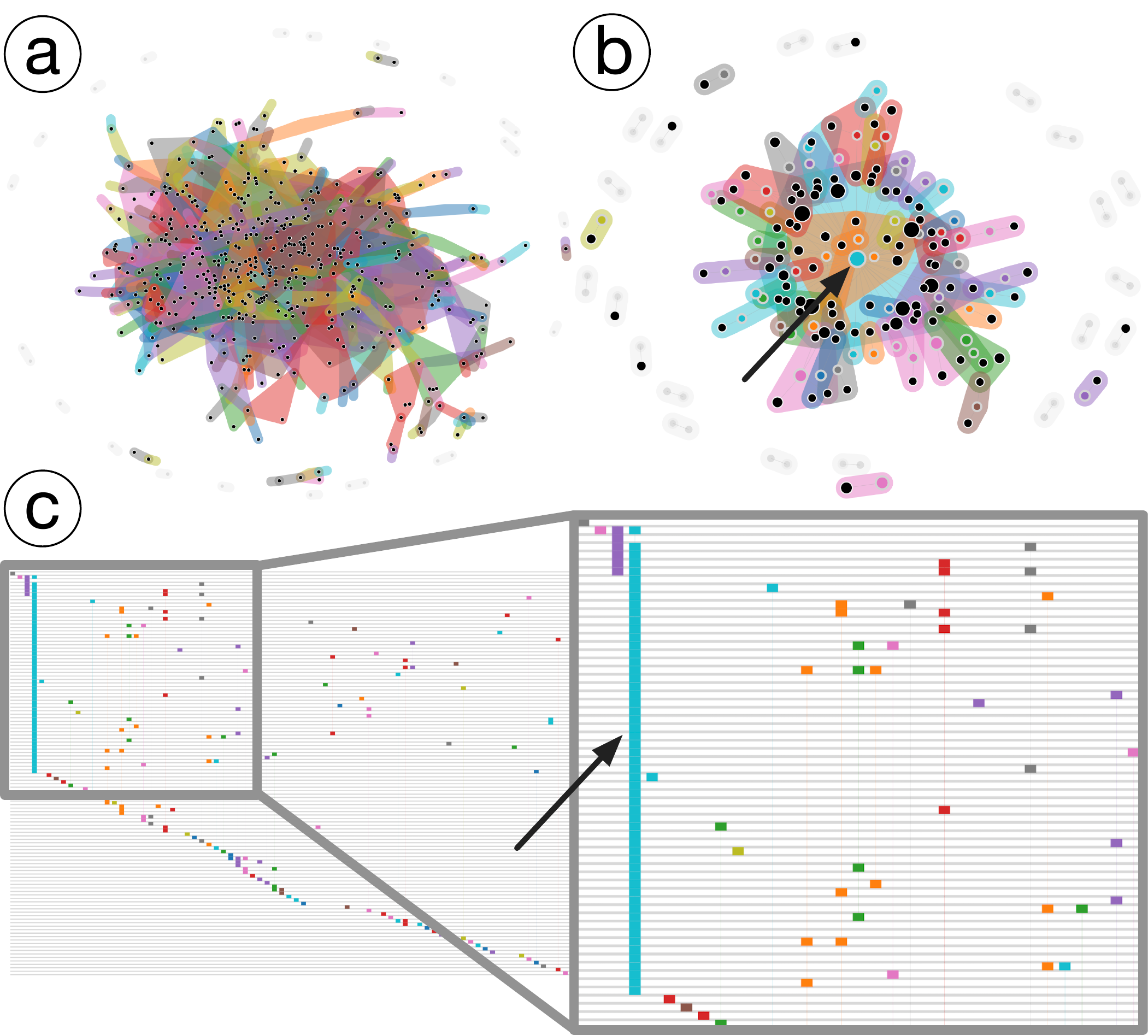}
    \vspace{-2mm}
    \caption{A hypergraph of the coauthor network from topic-coauthor dataset. (a) Original hypergraph. (b) A simplified hypergraph: the central hyperedge in blue (filled arrow) contains a number of prolific researchers with high number of papers and citation count. (c) The rainbow box based visualization of (b) highlighting  the researchers that belong to the blue hyperedge/column (filled arrow).}
    \label{fig:coauthor}
\end{figure}

Our final example concerns a coauthorship network in academic publications. 
We use the topic-coauthor dataset from ArnetMiner \cite{TangZhangYao2008} and focus specifically on the information-retrieval field-of-study for authors.
Each hyperedge represents an author and vertices in the corresponding hyperedge are researchers who have coauthored at least one paper with the author. 
There are 491 hyperedges and 1907 vertices in this hypergraph.
For this exploration, we use the following parameter setting: do not collapse edges or vertices, edge simplification, $s = 1$, grey out singletons, Jaccard weights, and $\epsilon = 5.5$.
\autoref{fig:coauthor}a shows the unsimplified hypergraph, and though there seems to be a central clustering affinity, it is very hard to parse the information in a hypergraph of this size.

Contrast this with the simplified hypergraph (at $s=1$, $\epsilon=5.5$) in \autoref{fig:coauthor}b, where the centrality of authorship is clearly evident. 
The central node (filled arrow) is the result of merging hyperedges under our framework.
The edges in this node correspond to prolific researchers in information retrieval with a high number of citations.
In particular, as illustrated by the rainbow box based visualization of the simplified hypergraph in \autoref{fig:coauthor}c, the central node consists of a hyperedge corresponding to \textbf{David A. Grossman} and \textbf{Ophir Frieder} who are the authors of the book ``Information Retrieval: Algorithms and Heuristics'', an important introductory textbook of the field.

%% file: sec-evaluation.tex
\begin{table*}[ht]
\begin{center}
\begin{tabular}{|c|c|c|c|c|c|c|c|c|c|c|c|c|}
    \hline
    {\multirow{2}{*}{Datasets}} & \multicolumn{3}{c|}{$m_i$} & \multicolumn{3}{c|}{$m_c$} & \multicolumn{3}{c|}{$m_l$} &  \multicolumn{3}{c|}{$m_a$}\\ 
    \cline{2-13}
    {} & B&A& B/A & B&A&B/A & B&A&B/A &  B&A&B/A\\ \hline
    Southern Women & 164&139&\textbf{1.18} & 0.94&0.95&\textbf{0.99}& 0.05&0.08&0.625& 0.32&0.46&\textbf{0.70} \\ \hline
    \lesmis & 1246&962&\textbf{1.30} & 0.98&0.94&1.04& 0.04&0.06&0.67& 0.56&0.54&1.04 \\ \hline
    Hallmarks Biological Pathways & 554&120&\textbf{4.62} & 0.99&0.92&1.08& 0.01&0.02&0.5& 0.88&0.48&1.83 \\ \hline
    ActiveDNS & 656&51&\textbf{12.86} & 0.96&0.93&1.03& 0.04&0.13&0.31& 0.45&0.58&\textbf{0.78} \\ \hline
    Coauthor Networks & 26162&278&\textbf{94.11} & 0.98&0.99&\textbf{0.99} & 0.02&0.05&0.4& 0.67&0.86&\textbf{0.78}  \\ \hline    
\end{tabular}
\end{center}
\vspace{-2mm}
\caption{Aesthetic metric values before (B) and after (A) simplification, and the ratio (R) of B to A. The simplification parameters are as follows. For \textit{southern women} dataset, vertex simplification, $s=1$, $\epsilon=1.6$. For \textit{\lesmis} dataset, vertex simplification, $s=1$, $\epsilon=2.93$. For \textit{Hallmarks biological pathways} dataset, edge simplification, $s=1$, $\epsilon=12.17$. For \textit{ActiveDNS} dataset, edge simplification, $s=3$, $\epsilon=2.4$. For \textit{coauthor networks} dataset, edge simplification, $s=1$, $\epsilon=5.5$. The bold numbers in ratio mean that the simplified visualization is better than the original visualization under the corresponding criterion.}
\label{tab:evaluation}
\end{table*}

\section{Evaluation}
\label{sec:evaluation}

It is important to note that the main contribution of this paper is the topology-based simplifications of hypergraphs, which can be used for \emph{any} hypergraph visualization technique. In other words, hypergraph simplification is considered \emph{orthogonal} to hypergraph visualization. 
On the other hand, to demonstrate the effectiveness of such simplifications, we evaluate the quality of hypergraph visualizations from~\autoref{sec:results} using four different aesthetic criteria detailed below. 

\para{Contour intersections.} Given a Venn diagram based visualization of a hypergraph $H$, a contour intersection is a crossing of the boundaries of two convex hulls representing two hyperedges. To compute the number of contour intersections, we approximate the boundary of each convex hull using piecewise linear segments. The \emph{approximated contour intersections}, denoted as $m_i$, is defined as the  number of intersections among these line segments. 

\para{Number of edge crossings.} Given a bipartite graph based visualization of a hypergraph $H$, we compute its \emph{number of edge crossings}, denoted as $m_c$, which is an aesthetic criterion first proposed by Purchase~\cite{Purchase2002}. 
For the bipartite graph representation of $H$ with vertex set $V$ and edge set $E$, $m_c$  is defined as
$$ 
m_c=\left\{
\begin{aligned}
1 - \frac{c}{c_{max}},~ \text{if } c_{max} > 0 \\
1,~ \text{otherwise}
\end{aligned}
\right.
$$
where $c$ is the number of edge crossings, and $c_{max}$ is the approximation of the upper bound of the number of edge crossings. $c_{max}$ is defined as
$$
c_{max} = \frac{|E|(|E|-1)}{2} - \frac{1}{2}\sum_{v\in V}(deg(v)(deg(v)-1))
$$
where $deg(v)$ is the degree of a vertex $v$.
Therefore, $0 \le m_c \le 1$, and $m_c = 1$ when there are no edge crossings in the bipartite graph.

\para{Normalized edge length variation.}
We also evaluate the bipartite graph visualization of a hypergraph using the \emph{normalized edge length variation}~\cite{KwonCrnovrsaninMa2017}, denoted as $m_l$. 
For a graph with a vertex set $V$ and an edge set $E$, Hachul and J\"{u}nger~\cite{HachulJunger2007} proposed a  \emph{normalized standard deviation of the edge length} $\sigma_{l}$, 
$$
\sigma_{l} = \sqrt{\frac{\sum_{e\in E}(l_e - l_{\mu})^2}{|E|\cdot l_{\mu}^2}}
$$
where $l_e$ is the length of an edge $e$, and $l_{\mu}$ is the mean of the edge length.
Kwon~\etal~\cite{KwonCrnovrsaninMa2017} then proposed a normalized version of $\sigma_{l}$, which is to divide $\sigma_{l}$  by its upper bound $\sqrt{|E|-1}$, and the normalized edge length variation $m_l$ is then defined as
$$
m_l = \frac{\sigma_l}{\sqrt{|E|-1}}. 
$$
By definition, $0 \le m_l \le 1$. $m_l = 0$ when all edge lengths are equal. 

\para{Minimum angle}. 
Finally, we work with the the \emph{minimum angle between adjacent edges leaving a node}~\cite{Purchase2002}, denoted as $m_a$. 
Given a vertex $v$, $m_a$ is defined based on the deviation of the minimum angle $\theta_{min}(v)$ between the adjacent incident edges from the ideal minimum angle $\theta(v)$, where $\theta(v) = \frac{360^{\circ}}{deg(v)}$. The average absolute deviation of minimum angles is defined as
$$
d_{\theta} = \frac{1}{|V|} \sum_{v \in V} \displaystyle\left\lvert \frac{\theta(v) - \theta_{min}(v)}{\theta(v)} \right\rvert, 
$$
and the minimum angle metric $m_a$ is defined as
$$
m_a = 1 - d_{\theta}.
$$
By definition, $0 \le m_a \le 1$. $m_a=1$ when all the nodes have equal angles between all adjacent incident edges. 

\para{Evaluation results.}
We compute the above four criteria to evaluate the resulting hypergraph visualizations in~\autoref{sec:results} that are either based on the Venn diagram or the bipartite graph. For each hypergraph, we compute a given criterion before and after simplification using the visualizations generated automatically by our tool without any modification. We also compute the ratio of the reported values before and after simplification for comparison. The last three criteria are computed using \textit{GLAM} ({https://github.com/VIDILabs/glam}). 

The evaluation results are shown in~\autoref{tab:evaluation}. 
As shown in the table, our hypergraph simplification greatly improves the approximated number of contour intersections ($m_i$), in particular, for large and complex hypergraphs (e.g., the Coauthor Networks). 
This is indicated by a large ratio $B/A$ that ranges between $1.2\times$ and $94\times$ in the table.  
Meanwhile, we obtain comparable numbers of edge crossings $m_c$ for the bipartite graph visualization before and after simplification. 
This is indicated by a ratio $\approx 1$ before and after simplification. 
On the other hand, a bipartite graph visualization of the simplified hypergraph shows slightly higher normalized edge length variation ($m_l$). However, we consider this to be  acceptable since all values are very close to $0$ (ranging from $0.01$ to $0.13$ across all datasets).  
Finally, we observe that we improve upon the minimum angle criterion $m_a$ for three out of five of the datasets with the simplification.

%% file: sec-discussion.tex
\section{Discussion}
\label{sec:discussion}

In this paper, we introduce a framework that supports topological simplification of hypergraphs via both vertex and edge simplifications. 
We exploit the duality between hypergraphs, line graphs, and clique expansions; and apply barcode-guided simplification of a hypergraph across multiple scales. 
We expect our framework to be applicable for general hypernetwork science (e.g. to be integrated with HyperNetX~\cite{HyperNetX}). 
There are several interesting venues for future research. 

We map a hypergraph to a metric space representation, where we use a shortest path metric in our current framework. 
There are other notions of metrics, in particular, resistance distance~\cite{KleinRandic1993}, commute time distance~\cite{FoussPirotteRenders2007}, and diffusion distance~\cite{CoifmanLafonLee2005}, which are applicable in our unifying framework (see~\autoref{sec:method-simplify}).  
These metrics can be particularly interesting as they capture structural or physical properties of the underlying data. 

As the barcode used to guide the simplification can be obtained by computing the minimum spanning tree, the simplification process is equivalent to applying a single-linkage clustering with a threshold. Some other clustering approaches, such as average-linkage clustering and complete-linkage clustering, or more complex methods using hypergraph Laplacians \cite{HayashiAksoyPark2020} or modularity \cite{KaminskiPoulinPraat2019}, might also be used to guide hypergraph simplifications. 
A comparison among these clustering approaches will be interesting. 
Finally, it will be interesting to explore hypergraph simplification using higher dimensional barcode.

%% file: HyperGraph-arXiv-main.bbl

%% file: sec-bio.tex
\vskip -2.2\baselineskip plus -1fil
\begin{IEEEbiography}[{
\vspace{-4mm}
\includegraphics[width=1in,height=1.25in, clip,keepaspectratio]{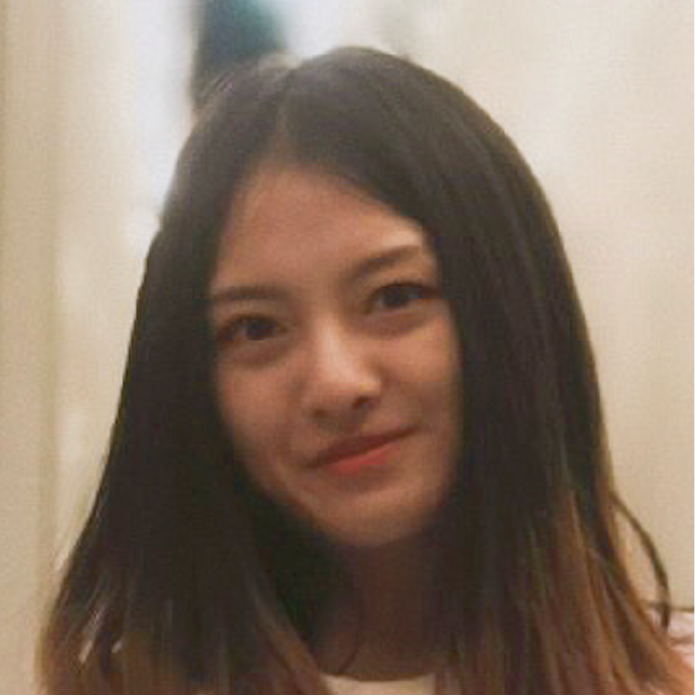}}]{Youjia Zhou}
is a PhD student at the School of Computing and the Scientific Computing and Imaging (SCI) Institute, University of Utah. 
Her research focuses on developing visual analytics systems for large and complex data, in particular, networks, high-dimensional point clouds, and vector/tensor fields, using topological techniques.
\end{IEEEbiography}
\vskip -2.2\baselineskip plus -1fil
\begin{IEEEbiography}[{
\vspace{-4mm}
\includegraphics[width=1in,height=1.25in, clip,keepaspectratio]{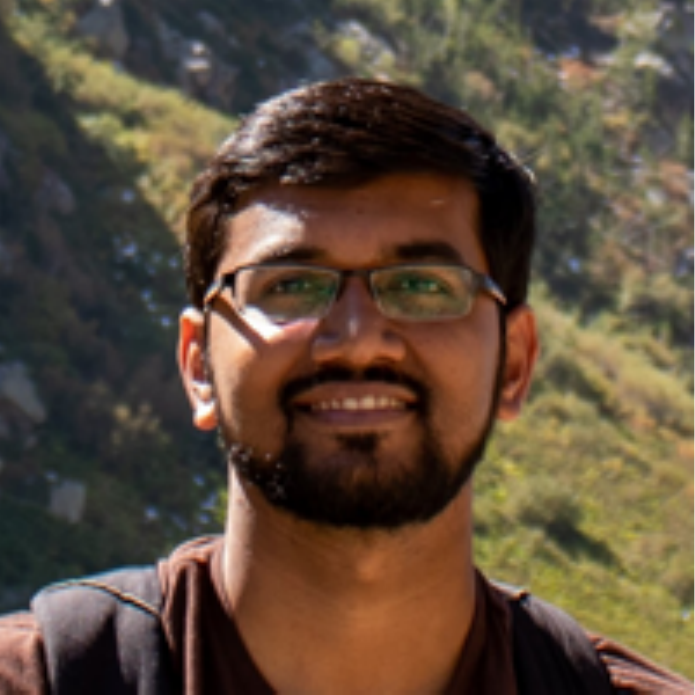}}]{Archit Rathore}
is a PhD student at the School of Computing and the Scientific Computing and Imaging (SCI) Institute, University of Utah.  His current research focuses on probing machine learning models through visualization techniques to improve interpretability.
\end{IEEEbiography}
\vskip -2.2\baselineskip plus -1fil
\begin{IEEEbiography}[{
\vspace{-4mm}
\includegraphics[width=1in,height=1.25in,clip,keepaspectratio]{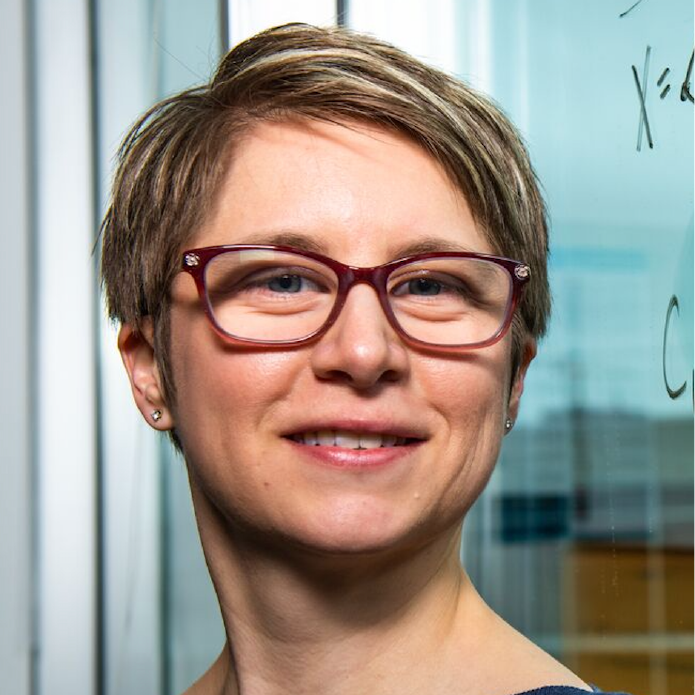}}]{Emilie Purvine} 
is a Senior Data Scientist at Pacific Northwest National Laboratory (PNNL). She received a Ph.D. in Mathematics from Rutgers University in May 2011 with a focus on experimental mathematics and nonlinear recurrence relations. At PNNL she is now focused on applications of combinatorics and computational topology together with theoretical advances needed to support the applications, ranging from computational chemistry and biology to cyber security and power grid modeling. 
\end{IEEEbiography}
\vskip -2.2\baselineskip plus -1fil
\begin{IEEEbiography}[{
\vspace{-4mm}
\includegraphics[width=1in,height=1.25in,clip,keepaspectratio]{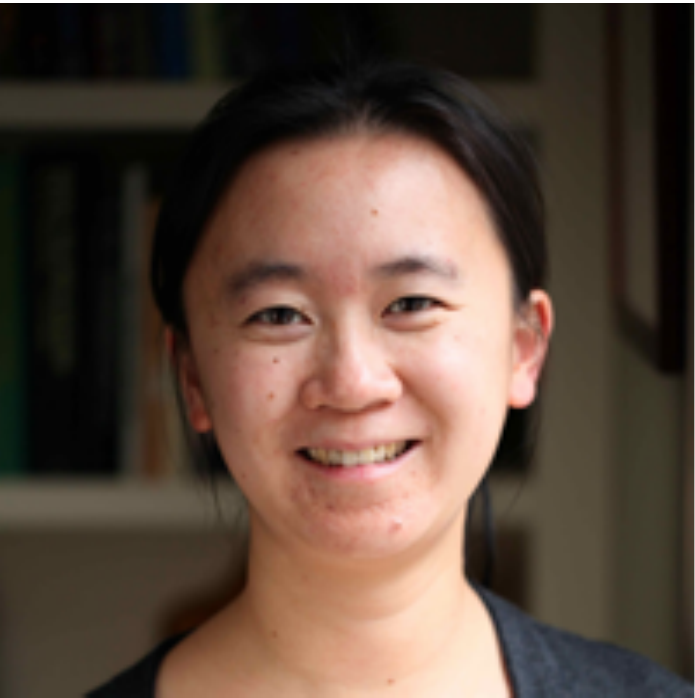}}]{Bei Wang} is an assistant professor at the School of Computing and
a faculty member at the Scientific Computing and Imaging (SCI)
Institute, University of Utah. She received her Ph.D. in Computer
Science from Duke University. She is interested in the analysis and
visualization of large and complex data. Her research interests
include topological data analysis, data visualization, computational
topology, computational geometry, machine learning and data mining.  
\end{IEEEbiography}

%% file: HyperGraph-arXiv-main.bbl
\begin{thebibliography}{10}
\providecommand{\url}[1]{#1}
\csname url@samestyle\endcsname
\providecommand{\newblock}{\relax}
\providecommand{\bibinfo}[2]{#2}
\providecommand{\BIBentrySTDinterwordspacing}{\spaceskip=0pt\relax}
\providecommand{\BIBentryALTinterwordstretchfactor}{4}
\providecommand{\BIBentryALTinterwordspacing}{\spaceskip=\fontdimen2\font plus
\BIBentryALTinterwordstretchfactor\fontdimen3\font minus
  \fontdimen4\font\relax}
\providecommand{\BIBforeignlanguage}[2]{{%
\expandafter\ifx\csname l@#1\endcsname\relax
\typeout{** WARNING: IEEEtran.bst: No hyphenation pattern has been}%
\typeout{** loaded for the language `#1'. Using the pattern for}%
\typeout{** the default language instead.}%
\else
\language=\csname l@#1\endcsname
\fi
#2}}
\providecommand{\BIBdecl}{\relax}
\BIBdecl

\bibitem{JoslynAksoyArendt2020}
C.~A. Joslyn, S.~Aksoy, D.~Arendt, J.~Firoz, L.~Jenkins, B.~Praggastis, E.~A.
  Purvine, and M.~Zalewski, ``Hypergraph analytics of domain name system
  relationships,'' in \emph{Procceedings of the 17th Workshop on Algorithms and
  Models for the Web Graph, Lecture Notes in Computer Science}, 2020.

\bibitem{AshburnerBallBlake2000}
M.~Ashburner, C.~A. Ball, J.~A. Blake, D.~Botstein, H.~Butler, J.~M. Cherry,
  A.~P. Davis, K.~Dolinski, S.~S. Dwight, J.~T. Eppig, M.~A. Harris, D.~P.
  Hill, L.~Issel-Tarver, A.~Kasarskis, S.~Lewis, J.~C. Matese, J.~E.
  Richardson, M.~Ringwald, G.~M. Rubin, and G.~Sherlock, ``Gene ontology: tool
  for the unification of biology,'' \emph{Nature Genetics}, vol.~25, no.~1, pp.
  25--9, 2000.

\bibitem{GeneOntologyConsortium2019}
{The Gene Ontology Consortium}, ``The gene ontology resource: 20 years and
  still {GOing} strong,'' \emph{Nucleic Acids Research}, vol.~47, no.~D1, pp.
  D330--D338, 2019.

\bibitem{LiberzonBirgerThorvaldsdottir2015}
A.~Liberzon, C.~Birger, H.~Thorvaldsd\'{o}ttir, M.~Ghandi, J.~P. Mesirov, and
  P.~Tamayo, ``The molecular signatures database hallmark gene set
  collection,'' \emph{Cell systems}, vol.~1, no.~6, pp. 417--425, 2015.

\bibitem{SubramanianTamayoMootha2005}
A.~Subramanian, P.~Tamayo, V.~K. Mootha, S.~Mukherjee, B.~L. Ebert, M.~A.
  Gillette, A.~Paulovich, S.~L. Pomeroy, T.~R. Golub, E.~S. Lander, and J.~P.
  Mesirov, ``Gene set enrichment analysis: A knowledge-based approach for
  interpreting genome-wide expression profiles,'' \emph{Proceedings of the
  National Academy of Sciences}, vol. 102, no.~43, pp. 15\,545--15\,550, 2005.

\bibitem{Cohen-SteinerEdelsbrunnerHarer2007}
D.~Cohen-Steiner, H.~Edelsbrunner, and J.~Harer, ``Stability of persistence
  diagrams,'' \emph{Discrete \& Computational Geometry}, vol.~37, pp. 103--120,
  2007.

\bibitem{HermanMelanconMarshall2000}
I.~Herman, G.~Melancon, and M.~S. Marshall, ``Graph visualization and
  navigation in information visualization: A survey,'' \emph{{IEEE}
  Transactions of Visualization and Computer Graphics}, vol.~6, no.~1, pp.
  24--43, 2000.

\bibitem{WangTao2017}
C.~Wang and J.~Tao, ``Graphs in scientific visualization: A survey,''
  \emph{Computer Graphic Forum}, vol.~36, no.~1, pp. 263--287, 2017.

\bibitem{LandesbergerKuijperSchreck2011}
T.~Von~Landesberger, A.~Kuijper, T.~Schreck, J.~Kohlhammer, J.~J. van Wijk,
  J.-D. Fekete, and D.~W. Fellner, ``Visual analysis of large graphs:
  State-of-the-art and future research challenges,'' \emph{Computer Graphics
  Forum}, vol.~30, no.~6, pp. 1719--1749, 2011.

\bibitem{BeckBurchDiehl2014}
F.~Beck, M.~Burch, S.~Diehl, and D.~Weiskopf, ``The state of the art in
  visualizing dynamic graphs,'' \emph{EuroVis STARs}, vol.~2, 2014.

\bibitem{EadesTamassia1994}
P.~Eades and R.~Tamassia, ``Algorithms for drawing graphs: An annotated
  bibliography,'' \emph{Computational Geometry: Theory and Applications},
  vol.~4, no.~5, pp. 235--282, 1994.

\bibitem{Makinen1990}
E.~M\"{a}kinen, ``How to draw a hypergraph,'' \emph{International Journal of
  Computer Mathematics}, vol.~34, no. 3-4, pp. 177--185, 1990.

\bibitem{ArafatBressan2017}
N.~A. Arafat and S.~Bressan, ``Hypergraph drawing by force-directed
  placement,'' in \emph{International Conference on Database and Expert Systems
  Applications}, 2017, pp. 387--394.

\bibitem{PaquetteTokuyasu2011}
J.~Paquette and T.~Tokuyasu, ``Hypergraph visualization and enrichment
  statistics: How the {EGAN} paradigm facilitates organic discovery from big
  data,'' in \emph{Human Vision and Electronic Imaging {XVI}}, vol. 7865, no.
  78650E.\hskip 1em plus 0.5em minus 0.4em\relax International Society for
  Optics and Photonics, 2011.

\bibitem{AlsallakhMicallefAigner2016}
B.~Alsallakh, L.~Micallef, W.~Aigner, H.~Hauser, S.~Miksch, and P.~Rodgers,
  ``The state-of-the-art of set visualization,'' \emph{Computer Graphics
  Forum}, vol.~35, no.~1, pp. 234--260, 2016.

\bibitem{Euler1761}
L.~Euler, ``Lettres \'{a} une princesse d'allemagne,'' \emph{{Letters
  102-105}}, 1761.

\bibitem{KritzPerlin1994}
M.~Kritz and K.~Perlin, ``A new scheme for drawing hypergraphs,''
  \emph{International journal of computer mathematics}, vol.~50, no. 3-4, pp.
  131--134, 1994.

\bibitem{SimonettoAuberArchambault2009}
P.~Simonetto, D.~Auber, and D.~Archambault, ``Fully automatic visualisation of
  overlapping sets,'' \emph{Computer Graphic Forum}, vol.~28, no.~3, pp.
  967--974, 2009.

\bibitem{Simonetto2011}
P.~Simonetto, ``Visualisation of overlapping sets and clusters with {Euler}
  diagrams,'' Ph.D. dissertation, University of Bordeaux, 2011.

\bibitem{AlperRicheRamos2011}
B.~Alper, N.~Riche, G.~Ramos, and M.~Czerwinski, ``Design study of linesets, a
  novel set visualization technique,'' \emph{IEEE transactions on visualization
  and computer graphics}, vol.~17, no.~12, pp. 2259--2267, 2011.

\bibitem{CollinsPennCarpendale2009}
C.~Collins, G.~Penn, and S.~Carpendale, ``Bubble sets: Revealing set relations
  with isocontours over existing visualizations,'' \emph{IEEE Transactions on
  Visualization and Computer Graphics}, vol.~15, no.~6, pp. 1009--1016, 2009.

\bibitem{EfratHuKobourov2015}
A.~Efrat, Y.~Hu, S.~G. Kobourov, and S.~Pupyrev, ``{MapSets}: Visualizing
  embedded and clustered graphs,'' \emph{Journal of Graph Algorithms and
  Applications}, vol.~19, no.~2, pp. 571--593, 2015.

\bibitem{LexGehlenborgStrobelt2014}
A.~Lex, N.~Gehlenborg, H.~Strobelt, R.~Vuillemot, and H.~Pfister, ``{UpSet}:
  visualization of intersecting sets,'' \emph{{IEEE} Transactions on
  Visualization and Computer Graphics}, vol.~20, no.~12, pp. 1983--1992, 2014.

\bibitem{RodgersStapletonChapman2015}
P.~Rodgers, G.~Stapleton, and P.~Chapman, ``Visualizing sets with linear
  diagrams,'' \emph{ACM Transactions on Computer-Human Interaction}, vol.~22,
  no.~6, pp. 1--39, 2015.

\bibitem{Lamy2019}
J.-B. Lamy, ``Visualizing undirected graphs and symmetric square matrices as
  overlapping sets,'' \emph{Multimedia Tools and Applications}, vol.~78,
  no.~23, pp. 33\,091--33\,112, 2019.

\bibitem{HyperNetX}
B.~Praggastis, D.~Arendt, E.~Purvine, C.~Joslyn, M.~Raugas, S.~Aksoy, and
  K.~Monson, ``{HyperNetX},'' \url{https://github.com/pnnl/HyperNetX}, 2018.

\bibitem{ValdiviaBuonoPlaisant2021}
P.~Valdivia, P.~Buono, C.~Plaisant, N.~Dufournaud, and J.-D. Fekete,
  ``Analyzing dynamic hypergraphs with {Parallel Aggregated Ordered Hypergraph}
  visualization,'' \emph{{IEEE} Transactions on Visualization and Computer
  Graphics}, vol.~27, no.~1, pp. 1--13, 2021.

\bibitem{KerrenJusufi2013}
A.~Kerren and I.~Jusufi, ``A novel radial visualization approach for undirected
  hypergraphs,'' in \emph{Proceedings of the 17th Eurographics Conference on
  Visualization, Short paper track}, 2013.

\bibitem{BrandesCornelsenPampel2012}
U.~Brandes, S.~Cornelsen, B.~Pampel, and A.~Sallaberry, ``Path-based supports
  for hypergraphs,'' \emph{Journal of Discrete Algorithms}, vol.~14, pp.
  248--261, 2012.

\bibitem{BuchinKreveldMeijer2009}
K.~Buchin, M.~van Kreveld, H.~Meijer, B.~Speckmann, and K.~Verbeek, ``On planar
  supports for hypergraphs,'' \emph{International Symposium on Graph Drawing},
  pp. 345--356, 2009.

\bibitem{Zykov1974}
A.~A. Zykov, ``Hypergraphs,'' \emph{Russian Mathematical Surveys}, vol.~29,
  no.~6, 1974.

\bibitem{Gropp1995}
H.~Gropp, ``The drawing of configurations,'' \emph{International Symposium on
  Graph Drawing}, pp. 267--276, 1995.

\bibitem{EvansRzazewskiSaeedi2019}
W.~Evans, P.~Rz\c{a}\.{z}ewski, N.~Saeedi, C.-S. Shin, and A.~Wolff,
  ``Representing graphs and hypergraphs by touching polygons in {3D},'' in
  \emph{International Symposium on Graph Drawing and Network Visualization},
  2019, pp. 18--32.

\bibitem{MeidianaHongEades2019}
A.~Meidiana, S.-H. Hong, P.~Eades, and D.~Keim, ``A quality metric for
  visualization of clusters in graphs,'' in \emph{International Symposium on
  Graph Drawing and Network Visualization}.\hskip 1em plus 0.5em minus
  0.4em\relax Springer, 2019, pp. 125--138.

\bibitem{NguyenEadesHong2012}
Q.~Nguyen, P.~Eades, and S.-H. Hong, ``On the faithfulness of graph
  visualizations,'' in \emph{International Symposium on Graph Drawing}.\hskip
  1em plus 0.5em minus 0.4em\relax Springer, 2012, pp. 566--568.

\bibitem{Purchase2002}
H.~C. Purchase, ``Metrics for graph drawing aesthetics,'' \emph{Journal of
  Visual Languages \& Computing}, vol.~13, no.~5, pp. 501--516, 2002.

\bibitem{KwonCrnovrsaninMa2017}
O.-H. Kwon, T.~Crnovrsanin, and K.-L. Ma, ``What would a graph look like in
  this layout? a machine learning approach to large graph visualization,''
  \emph{IEEE transactions on visualization and computer graphics}, vol.~24,
  no.~1, pp. 478--488, 2017.

\bibitem{purohit2014fast}
M.~Purohit, B.~A. Prakash, C.~Kang, Y.~Zhang, and V.~Subrahmanian, ``Fast
  influence-based coarsening for large networks,'' in \emph{Proceedings of the
  20th ACM SIGKDD international conference on Knowledge discovery and data
  mining}, 2014, pp. 1296--1305.

\bibitem{ShinGhotingKim2019}
K.~Shin, A.~Ghoting, M.~Kim, and H.~Raghavan, ``{SWeG}: Lossless and lossy
  summarization of web-scale graphs,'' in \emph{The World Wide Web Conference},
  2019, pp. 1679--1690.

\bibitem{beg2018scalable}
M.~A. Beg, M.~Ahmad, A.~Zaman, and I.~Khan, ``Scalable approximation algorithm
  for graph summarization,'' in \emph{Pacific-Asia Conference on Knowledge
  Discovery and Data Mining}.\hskip 1em plus 0.5em minus 0.4em\relax Springer,
  2018, pp. 502--514.

\bibitem{ShenMaEliassiRad2006}
Z.~Shen, K.-L. Ma, and T.~Eliassi-Rad, ``Visual analysis of large heterogeneous
  social networks by semantic and structural abstraction,'' \emph{{IEEE}
  Transactions on Visualization and Computer Graphics}, vol.~12, no.~6, pp.
  1427--1439, 2006.

\bibitem{LeeJoKo2020}
K.~Lee, H.~Jo, J.~Ko, S.~Lim, and K.~Shin, ``{SSumM}: Sparse summarization of
  massive graphs,'' \emph{Proceedings of the 26th ACM SIGKDD International
  Conference on Knowledge Discovery \& Data Mining}, pp. 144--154, 2020.

\bibitem{KoutraKangVreeken2014}
D.~Koutra, U.~Kang, J.~Vreeken, and C.~Faloutsos, ``{VoG}: Summarizing and
  understanding large graphs,'' \emph{Proceedings of the 2014 SIAM
  international conference on data mining}, pp. 91--99, 2014.

\bibitem{ShahKoutraZou2015}
N.~Shah, D.~Koutra, T.~Zou, B.~Gallagher, and C.~Faloutsos, ``{TimeCrunch}:
  Interpretable dynamic graph summarization,'' \emph{Proceedings of the 21th
  {ACM SIGKDD} International Conference on Knowledge Discovery and Data
  Mining}, pp. 1055--1064, 2015.

\bibitem{DunneShneiderman2013}
C.~Dunne and B.~Shneiderman, ``Motif simplification: improving network
  visualization readability with fan, connector, and clique glyphs,''
  \emph{Proceedings of the {SIGCHI} Conference on Human Factors in Computing
  Systems}, pp. 3247--3256, 2013.

\bibitem{SuhHajijWang2019}
A.~Suh, M.~Hajij, B.~Wang, C.~Scheidegger, and P.~Rosen, ``Persistent homology
  guided force-directed graph layouts,'' \emph{IEEE Transactions on
  Visualization and Computer Graphics}, vol.~26, no.~1, pp. 697--707, 2020.

\bibitem{LemonnierWeteringKissinger2020}
L.~Lemonnier, J.~van~de Wetering, and A.~Kissinger, ``Hypergraph
  simplification: Linking the path-sum approach to the {ZH-calculus},''
  \emph{arXiv preprint arXiv:2003.13564}, 2020.

\bibitem{aksoy2020hypernetwork}
S.~G. Aksoy, C.~Joslyn, C.~O. Marrero, B.~Praggastis, and E.~Purvine,
  ``Hypernetwork science via high-order hypergraph walks,'' \emph{EPJ Data
  Science}, vol.~9, no.~1, p.~16, 2020.

\bibitem{BermondHeydemannSotteau1977}
J.-C. Bermond, M.-C. Heydemann, and D.~Sotteau, ``Line graphs of hypergraphs
  {I},'' \emph{Discrete Mathematics}, vol.~18, no.~3, pp. 235--241, 1977.

\bibitem{ZienSchlagChan1999}
J.~Zien, M.~Schlag, and P.~K. Chan, ``Multi-level spectral hypergraph
  partitioning with arbitrary vertex sizes,'' \emph{{IEEE} Transactions on
  Computer-Aided Designof Integrated Circuits and Systems}, vol.~18, pp.
  1389--1399, 1999.

\bibitem{CarlssonZomorodianCollins2004}
G.~Carlsson, A.~J. Zomorodian, A.~Collins, and L.~J. Guibas, ``Persistence
  barcodes for shapes,'' \emph{Proceedings Eurographs/{ACM} {SIGGRAPH}
  Symposium on Geometry Processing}, pp. 124--135, 2004.

\bibitem{Ghrist2008}
R.~Ghrist, ``Barcodes: The persistent topology of data,'' \emph{Bullentin of
  the American Mathematical Society}, vol.~45, pp. 61--75, 2008.

\bibitem{EdelsbrunnerLetscherZomorodian2002}
H.~Edelsbrunner, D.~Letscher, and A.~Zomorodian, ``Topological persistence and
  simplification,'' \emph{Discrete and Computational Geometry}, vol.~28, pp.
  511--533, 2002.

\bibitem{HajijWangScheidegger2018}
M.~Hajij, B.~Wang, C.~Scheidegger, and P.~Rosen, ``Visual detection of
  structural changes in time-varying graphs using persistent homology,''
  \emph{IEEE Pacific Visualization Symposium}, 2018.

\bibitem{EdelsbrunnerHarer2008}
H.~Edelsbrunner and J.~Harer, ``Persistent homology - a survey,''
  \emph{Contemporary Mathematics}, vol. 453, pp. 257--282, 2008.

\bibitem{GowerRoss1969}
J.~C. Gower and G.~J. Ross, ``Minimum spanning trees and single linkage cluster
  analysis,'' \emph{Journal of the Royal Statistical Society: Series C (Applied
  Statistics)}, vol.~18, no.~1, pp. 54--64, 1969.

\bibitem{GerberBremerPascucci2010}
S.~Gerber, P.-T. Bremer, V.~Pascucci, and R.~Whitaker, ``Visual exploration of
  high dimensional scalar functions,'' \emph{{IEEE} Transactions on
  Visualization and Computer Graphics}, vol.~16, pp. 1271--1280, 2010.

\bibitem{DavisGardnerGardner2009}
A.~Davis, B.~B. Gardner, and M.~R. Gardner, \emph{Deep South: A social
  anthropological study of caste and class}.\hskip 1em plus 0.5em minus
  0.4em\relax University of South Carolina Press, 2009.

\bibitem{Freeman2003}
L.~C. Freeman, ``Finding social groups: A meta-analysis of the southern women
  data,'' http://moreno.ss.uci.edu/86.pdf, 2003.

\bibitem{Knuth1994}
\BIBentryALTinterwordspacing
D.~E. Knuth, \emph{The Stanford {GraphBase}: a platform for combinatorial
  computing}.\hskip 1em plus 0.5em minus 0.4em\relax New York: ACM Press and
  Addison-Wesley Publishing Company, 1994. [Online]. Available:
  \url{https://www-cs-faculty.stanford.edu/~knuth/sgb.html}
\BIBentrySTDinterwordspacing

\bibitem{kountouras2016enabling}
A.~Kountouras, P.~Kintis, C.~Lever, Y.~Chen, Y.~Nadji, D.~Dagon,
  M.~Antonakakis, and R.~Joffe, ``Enabling network security through active dns
  datasets,'' in \emph{International Symposium on Research in Attacks,
  Intrusions, and Defenses}.\hskip 1em plus 0.5em minus 0.4em\relax Springer,
  2016, pp. 188--208.

\bibitem{CHGL}
S.~Aksoy, S.~Harun, L.~Jenkins, C.~Joslyn, C.~Lightsey, H.~Medal, D.~Mentgen,
  T.~Stavenger, T.~Bhuiyan, and M.~Zalewski, ``{Chapel Hypergraph Library},''
  \url{https://github.com/pnnl/CHGL}, 2018.

\bibitem{JeLBhT18}
L.~P. Jenkins, T.~Bhuiyan, S.~Harun, C.~Lightsey, S.~Aksoy, T.~Stavenger,
  M.~Zalewski, H.~Medal, and C.~Joslyn, ``{Chapel Hypergraph Library (CHGL)},''
  in \emph{{IEEE} High Performance Extreme Computing Conference}, 2018.

\bibitem{BGP}
``{Hurricane Electric BGP Toolkit},'' \url{https://bgp.he.net/}.

\bibitem{TangZhangYao2008}
J.~Tang, J.~Zhang, L.~Yao, J.~Li, L.~Zhang, and Z.~Su, ``{ArnetMiner}:
  Extraction and miningof academic social networks,'' in \emph{Proceedings of
  the 14th ACM SIGKDD international conference on Knowledge discovery and data
  mining}, 2008, pp. 990--998.

\bibitem{HachulJunger2007}
S.~Hachul and M.~J{\"u}nger, ``Large-graph layout algorithms at work: An
  experimental study,'' \emph{Journal of Graph Algorithms and Applications},
  vol.~11, no.~2, pp. 345--369, 2007.

\bibitem{KleinRandic1993}
D.~J. Klein and M.~Randic, ``Resistance distance,'' \emph{Journal of
  Mathematical Chemistry}, vol.~12, pp. 81--95, 1993.

\bibitem{FoussPirotteRenders2007}
F.~Fouss, A.~Pirotte, J.~michel Renders, and M.~Saerens, ``Random-walk
  computation of similarities between nodes of a graph, with application to
  collaborative recommendation,'' \emph{{IEEE} Transactions on Knowledge and
  Data Engineering}, vol.~19, no.~3, pp. 355--369, 2007.

\bibitem{CoifmanLafonLee2005}
R.~R. Coifman, S.~Lafon, A.~B. Lee, M.~Maggioni, B.~Nadler, F.~Warner, and
  S.~W. Zucker, ``Geometric diffusions as a tool for harmonic analysisand
  structure definition of data: Diffusion maps,'' \emph{Proceedings of the
  National Academy of Sciences of the United States of America (PNAS)}, vol.
  102, no.~21, pp. 7426--7431, 2005.

\bibitem{HayashiAksoyPark2020}
K.~Hayashi, S.~G. Aksoy, C.~H. Park, and H.~Park, ``Hypergraph random walks,
  {Laplacians}, and clustering,'' in \emph{Proceedings of the 29th ACM
  International Conference on Information \& Knowledge Management}, 2020, pp.
  495--504.

\bibitem{KaminskiPoulinPraat2019}
B.~Kami{\'n}ski, V.~Poulin, P.~Pra{\l}at, P.~Szufel, and F.~Th{\'e}berge,
  ``Clustering via hypergraph modularity,'' \emph{PloS one}, vol.~14, no.~11,
  p. e0224307, 2019.

\end{thebibliography}
